\documentclass[12pt]{article}
\usepackage{amsmath,amssymb}
\usepackage{graphicx}
\newcommand{\eqdef}{\stackrel{\text{def}}{=}}
\newcommand{\n}{\nonumber\\}
\newcommand{\bm}{\boldsymbol}
\newcommand{\cF}{c_{\text{\tiny$\mathcal{F}$}}}
\newcommand{\ignore}[1]{}
\numberwithin{equation}{section}
\newcommand{\Romannumeral}[1]{\uppercase\expandafter{\romannumeral#1}}
\newcommand{\I}{\text{\Romannumeral{1}}}
\newcommand{\II}{\text{\Romannumeral{2}}}
\newcommand{\III}{\text{\Romannumeral{3}}}
\newtheorem{prop}{\bf Proposition}[section]

\allowdisplaybreaks[4]

\begin{document}

\baselineskip=20pt

\newfont{\elevenmib}{cmmib10 scaled\magstep1}
\newcommand{\preprint}{
    \begin{flushleft}
     \elevenmib Yukawa\, Institute\, Kyoto\\
   \end{flushleft}\vspace{-1.3cm}
   \begin{flushright}\normalsize \sf
     DPSU-12-3\\
     YITP-12-85\\
   \end{flushright}}
\newcommand{\Title}[1]{{\baselineskip=26pt
   \begin{center} \Large \bf #1 \\ \ \\ \end{center}}}
\newcommand{\Author}{\begin{center}
   \large \bf Satoru Odake${}^a$ and Ryu Sasaki${}^b$ \end{center}}
\newcommand{\Address}{\begin{center}
     $^a$ Department of Physics, Shinshu University,\\
     Matsumoto 390-8621, Japan\\
     ${}^b$ Yukawa Institute for Theoretical Physics,\\
     Kyoto University, Kyoto 606-8502, Japan\\
     E-mail:\ ryu@yukawa.kyoto-u.ac.jp
   \end{center}}
\newcommand{\Accepted}[1]{\begin{center}
   {\large \sf #1}\\ \vspace{1mm}{\small \sf Accepted for Publication}
   \end{center}}

\preprint
\thispagestyle{empty}

\Title{Krein-Adler transformations for shape-invariant potentials
and pseudo virtual states}

\Author

\Address
\vspace{1cm}

\begin{abstract}
For eleven examples of one-dimensional quantum mechanics with shape-invariant
potentials, the Darboux-Crum transformations in terms of multiple
{\em pseudo virtual state wavefunctions} are shown to be equivalent to
Krein-Adler transformations deleting multiple eigenstates with
{\em shifted parameters}.
These are based upon infinitely many polynomial Wronskian identities of
classical orthogonal polynomials, {\em i.e.} the Hermite, Laguerre and
Jacobi polynomials, which constitute the main part of the eigenfunctions
of various quantum mechanical systems with shape-invariant potentials.
\end{abstract}

\section{Introduction}
\label{intro}

The virtual state wavefunctions are essential for the construction of the
multi-indexed Laguerre and Jacobi polynomials \cite{os25,gomez3}.
They are polynomial type solutions of one-dimensional Schr\"odinger
equations for shape-invariant potentials \cite{genden,infhul,susyqm}.
They are characterised as having negative energies (the groundstate has
zero energy), no zeros in the physical domain and that they and their
reciprocals are square non-integrable.
By dropping the condition of no zeros and the reciprocals are required to be
square-integrable at both boundaries \eqref{type3},
{\em pseudo virtual state wavefunctions} are obtained.
In most cases, the virtual and pseudo virtual state wavefunctions are
obtained from the eigenfunctions by twisting the parameter(s) based on
the discrete symmetries of the Hamiltonian \cite{os25}.
Starting from a shape-invariant potential, a Darboux transformation
\cite{darb,crum} in terms of a nodeless pseudo virtual state wavefunction
$\tilde{\varphi}(x)$ with energy $\tilde{\mathcal{E}}$ produces a solvable
system with an extra eigenstate below the original groundstate with energy
$\tilde{\mathcal{E}}$ and eigenfunction $\tilde{\varphi}(x)^{-1}$.
This  method of generating a solvable system by ``adding an eigenstate''
below the groundstate is known for many years, starting from the simplest
harmonic oscillator potential examples \cite{dubov} and followed by many
authors \cite{gomez6}--\cite{quesne5}.
As remarked by Adler \cite{adler} for the harmonic oscillator case and
generalised by the present authors \cite{gos} for other potentials, such
a system can be derived by special types of Krein-Adler transformations.
That is, the Krein-Adler transformation for a system with negatively
shifted parameters in which the created state will be the groundstate.
The transformation use all the eigenstates between the new and the original
groundstates.

In this paper we present straightforward generalisation of the above
result for various shape-invariant potentials listed in section
\ref{sec:Exa}; Coulomb potential with the centrifugal barrier (C),
Kepler problem in spherical space (K), Morse potential (M),
soliton potential (s), Rosen-Morse potential (RM), hyperbolic symmetric
top $\II$ (hst), Kepler problem in hyperbolic space (Kh),
hyperbolic Darboux-P\"{o}schl-Teller potential (hDPT), on top of the
well-known harmonic oscillator (H), the radial oscillator (L) and
the Darboux-P\"oschl-Teller potential (J).
They are divided into two groups according to the eigenfunction
patterns in \S\,\ref{sec:2group}.
We mainly follow Infeld-Hull \cite{infhul} for the naming of potentials.
A Darboux-Crum transformation in terms of multiple pseudo virtual state
wavefunctions is equivalent to a certain Krein-Adler transformation
deleting multiple eigenstates with shifted parameters.
In contrast to the use of genuine virtual state wavefunctions \cite{os25},
not all choices of the multiple pseudo virtual states would generate
singularity free systems.
The singularity free conditions of the obtained system are supplied by
the known ones for the Krein-Adler transformations \cite{adler}.

Underlying the above equivalence are infinitely many polynomial Wronskian
identities relating Wronskians of polynomials with twisted parameters to
those of shifted parameters. These identities imply the equality of the
deformed potentials with the twisted and shifted parameters.
This in turn guarantees the equivalence of all the other eigenstate
wavefunctions.
We present the polynomial Wronskian identities for Group A;
the harmonic oscillator (H), the radial oscillator (L) and the
Darboux-P\"oschl-Teller potential (J) and some others.
For Group B, the identities take slightly different forms;
determinants of various polynomials with twisted and shifted parameters.
The infinitely many polynomial Wronskian identities are the consequences
of the fundamental Wronskian (determinant) identity \eqref{Wformula2} as
demonstrated in section \ref{sec:main}.

This paper is organised as follows.
The essence of Darboux-Crum transformations for the Schr\"odinger equation
in one dimension is recapitulated in \S\,\ref{sec:genstr}.
The definitions of virtual states and pseudo virtual states are given in
\S\,\ref{sec:vir}.
In section \ref{sec:Exa} two groups of eigenfunction patterns are
introduced in \S\,\ref{sec:2group} and related Wronskian expressions are
explored in \S\,\ref{sec:Wro}. The details of the eleven examples of
shape-invariant systems are provided in \S\,\ref{sec:H}--\S\,\ref{sec:hDPT}.
Section \ref{sec:main} is the main part of the paper.
We demonstrate the equivalence of the Darboux-Crum transformations
in terms of multiple pseudo virtual states
to Krein-Adler transformations in terms of multiple eigenstates with
shifted parameters. The underlying polynomial Wronskian identities are
proven with their more general determinant identities.
The final section is for a summary and comments. 

\section{Darboux-Crum \& Krein-Adler Transformations}
\label{sec:darb-crum}

Darboux transformations in general \cite{darb} apply to generic second
order differential equations of Schr\"odinger form
\begin{equation}
  \mathcal{H}=-\frac{d^2}{dx^2}+U(x),\quad
  \mathcal{H}\psi(x)=\mathcal{E}\psi(x)\ \quad
  \bigl(\mathcal{E},U(x)\in\mathbb{C}\bigr),
  \label{schr}
\end{equation}
without further structures of quantum mechanics, {\em e.g.} the boundary
conditions, self-adjointness of $\mathcal{H}$, Hilbert space, etc.
In the next subsection, we summarise the formulas of multiple Darboux
transformations, which are purely algebraic.

\subsection{General structure}
\label{sec:genstr}

Let $\{\varphi_j(x), \tilde{\mathcal{E}}_j\}$ ($j=1,2,\ldots,M$) be distinct
solutions of the original Schr\"odinger equation \eqref{schr}:
\begin{equation}
  \mathcal{H}\varphi_j(x)=\tilde{\mathcal{E}}_j\varphi_j(x)\quad
  (\tilde{\mathcal{E}}_j\in\mathbb{C}\ ;\ j=1,2,\ldots,M),
  \label{scheq2}
\end{equation}
to be called {\em seed} solutions.
By picking up one of the above seed solutions, say  $\varphi_1(x)$,
we form new functions with the above solution $\psi(x)$ and the rest
of $\{\varphi_k(x),\tilde{\mathcal{E}}_k\}$ ($k\neq 1$):
\begin{equation}
  \psi^{[1]}(x)\eqdef\frac{\text{W}[\varphi_1,\psi](x)}{\varphi_1(x)}
  =\frac{\varphi_1(x)\partial_x\psi(x)
  -\partial_x\varphi_1(x)\psi(x)}{\varphi_1(x)},\quad
  \varphi^{[1]}_{1,k}(x)\eqdef
  \frac{\text{W}[\varphi_1,\varphi_k](x)}{\varphi_1(x)}.
\end{equation}
It is elementary to show that $\psi^{[1]}(x)$,
$\varphi_1^{-1}(x)\,\bigl(\eqdef\varphi_1(x)^{-1}\bigr)$ and
$\varphi^{[1]}_{1,k}(x)$ are solutions of a new Schr\"odinger equation
of a deformed Hamiltonian $\mathcal{H}^{[1]}$ 
\begin{equation}
  \mathcal{H}^{[1]}=-\frac{d^2}{dx^2}+U^{[1]}(x),\quad
  U^{[1]}(x)\eqdef U(x)-2\partial_x^2\log\bigl|\varphi_1(x)\bigr|,
  \label{1pot}
\end{equation}
with the same energies $\mathcal{E}$, $\tilde{\mathcal{E}}_1$ and
$\tilde{\mathcal{E}}_k$:
\begin{align}
  \mathcal{H}^{[1]}\psi^{[1]}(x)&=\mathcal{E}\psi^{[1]}(x),\quad 
  \mathcal{H}^{[1]}\varphi^{-1}_1(x)=\tilde{\mathcal{E}}_1\varphi^{-1}_1(x),
  \label{newschr}\\
  \mathcal{H}^{[1]}\varphi^{[1]}_{1,k}(x)
  &=\tilde{\mathcal{E}}_k\varphi^{[1]}_{1,k}(x)\ \ (k\neq 1).
  \label{newschr2}
\end{align}
By repeating the above Darboux transformation $M$-times, we obtain new
functions
\begin{align}
  \psi^{[M]}(x)&\eqdef
  \frac{\text{W}[\varphi_1,\varphi_2,\ldots,\varphi_M,\psi](x)}
  {\text{W}[\varphi_1,\varphi_2,\ldots,\varphi_M](x)},
  \label{psiM}\\
  \breve{\varphi}^{[M]}_j(x)&\eqdef
  \frac{\text{W}[\varphi_1,\varphi_2,\ldots,\breve{\varphi}_j,\ldots,
  \varphi_M](x)}
  {\text{W}[\varphi_1,\varphi_2,\ldots,\varphi_M](x)}\ \ (j=1,2,\ldots,M),
  \label{varphiM}
\end{align}
which satisfy an $M$-th deformed Schr\"odinger equation with the energies
$\mathcal{E}$ and $\tilde{\mathcal{E}}_j$ \cite{crum,adler}:
\begin{align}
  &\mathcal{H}^{[M]}=-\frac{d^2}{dx^2}+U^{[M]}(x),\quad
  U^{[M]}(x)\eqdef U(x)-2\partial_x^2\log\bigl|
  \text{W}[\varphi_1,\varphi_2,\ldots,\varphi_M](x)\bigr|,
  \label{Mpot}\\
  &\mathcal{H}^{[M]}\psi^{[M]}(x)=\mathcal{E}\psi^{[M]}(x),\quad
  \mathcal{H}^{[M]}\breve{\varphi}^{[M]}_j(x)
  =\tilde{\mathcal{E}}_j\breve{\varphi}^{[M]}_j(x)\ \ (j=1,2,\ldots,M).
  \label{Mschr}
\end{align}
Here $\text{W}[f_1,f_2,\ldots,f_n](x)$ is a Wronskian
\begin{equation*}
  \text{W}[f_1,f_2,\ldots,f_n](x)\eqdef
  \det\Bigl(\frac{d^{j-1}f_k(x)}{dx^{j-1}}\Bigr)_{1\le j,k\le n}.
\end{equation*}
For $n=0$, we set $\text{W}[\cdot](x)=1$ and
$\text{W}[f_1,f_2,\ldots,\breve{f}_j,\ldots,f_n](x)$ means that $f_j(x)$ is
excluded from the Wronskian.
In deriving the determinant formulas \eqref{psiM}--\eqref{varphiM} and
\eqref{Mpot} use is made of the properties of the Wronskian
\begin{align}
  &\text{W}[gf_1,gf_2,\ldots,gf_n](x)
  =g(x)^n\text{W}[f_1,f_2,\ldots,f_n](x),
  \label{Wformula1}\\
  &\text{W}\bigl[\text{W}[f_1,f_2,\ldots,f_n,g],
  \text{W}[f_1,f_2,\ldots,f_n,h]\,\bigr](x)\n
  &=\text{W}[f_1,f_2,\ldots,f_n](x)\cdot
  \text{W}[f_1,f_2,\ldots,f_n,g,h](x)
  \qquad(n\geq 0).
  \label{Wformula2}
\end{align}
Another useful property of the Wronskian is that it is invariant when
the derivative $\frac{d}{dx}$ is replaced by an arbitrary `covariant
derivative' $D_i$ with an arbitrary smooth function $q_i(x)$:
\begin{equation}
  D_i\eqdef\frac{d}{dx}-q_i(x),\quad
  \text{W}[f_1,f_2,\ldots,f_n](x)=
  \det\bigl(D_{j-1}\cdots D_2D_1f_k(x)\bigr)_{1\le j,k\le n},
  \label{covW}
\end{equation}
with $D_{j-1}\cdots D_2D_1\bigm|_{j=1}=1$.
Under the change of variable $x\to\eta(x)$, the Wronskian behaves
\begin{equation}
  f_j(x)=F_j\bigl(\eta(x)\bigr),\quad
  \text{W}[f_1,f_2,\ldots,f_n](x)
  =\Bigl(\frac{d\eta(x)}{dx}\Bigr)^{\frac12n(n-1)}
  \text{W}[F_1,F_2,\ldots,F_n]\bigl(\eta(x)\bigr).
  \label{Wformula3}
\end{equation}

Obviously the potential $U^{[M]}(x)$ \eqref{Mpot} is independent of the
order of the seed solutions.
The zeros of the seed solution $\varphi_1(x)$ (the Wronskian
$\text{W}[\varphi_1,\ldots,\varphi_M](x)$) induce singularities of
the potential $U^{[1]}(x)$ in \eqref{1pot} ($U^{[M]}(x)$ in \eqref{Mpot}).

We apply Darboux transformations for various extensions of exactly solvable
potentials $U(x)\in\mathbb{R}$ in one dimensional quantum mechanics defined
in an interval $x_1<x<x_2$.
We assume that the potential is smooth in the interval and the system
has a finite (or an infinite) number of discrete eigenstates with a
vanishing groundstate energy:
\begin{align}
  \mathcal{H}\phi_n(x)&=\mathcal{E}_n\phi_n(x)\quad
  (\,0\le n\le n_{\text{max}}\ \ \text{or}\ \ n\in\mathbb{Z}_{\ge0}),
  \label{sheq}\\
  0&=\mathcal{E}_0<\mathcal{E}_1<\mathcal{E}_2<\cdots,\n
  (\phi_m,\phi_n)&\eqdef\int_{x_1}^{x_2}\!dx\,\phi_m(x)\phi_n(x)
  =h_n\delta_{m\,n},\quad h_n>0.
  \label{inpro}
\end{align}
Hereafter we consider real solutions only and use the term {\em eigenstates}
in their strict sense, {\em i.e.} they only apply to those with square
integrable wavefunctions and correspondingly the {\em eigenvalues}.
The zero groundstate energy condition can always be achieved by adjusting
the constant part of the potential $U(x)$.
Then the Hamiltonian is positive semi-definite and it has a simple
factorised form expressed in terms of the groundstate wavefunction
$\phi_0(x)$ which has no node in $(x_1,x_2)$:
\begin{align}
  \mathcal{H}&=-\frac{d^2}{dx^2}+U(x)
  =\mathcal{A}^{\dagger}\mathcal{A},\quad
  U(x)=\frac{\partial_x^2\phi_0(x)}{\phi_0(x)}
  =\bigl(\partial_x w(x)\bigr)^2+\partial_x^2w(x),\\
  \mathcal{A}&\eqdef\frac{d}{dx}-\partial_xw(x),\quad
  \mathcal{A}^{\dagger}=-\frac{d}{dx}-\partial_xw(x),\quad
  w(x)\in\mathbb{R},\quad\phi_0(x)=e^{w(x)}.
\end{align}
For all the examples of solvable potentials to be discussed in this paper,
the main part of the discrete eigenfunctions $\{\phi_n(x)\}$ are
{\em polynomials} $P_n(\eta(x))$, {\em in a certain function} $\eta(x)$,
which is called the {\em sinusoidal coordinate} \cite{os7}.
The original systems to be extended usually contain some parameter(s),
$\bm{\lambda}=(\lambda_1,\lambda_2,\ldots)$ and the parameter dependence
is denoted by $\mathcal{H}(\bm{\lambda})$,
$\mathcal{A}(\bm{\lambda})$, $\mathcal{E}_n(\bm{\lambda})$,
$\phi_n(x;\bm{\lambda})$, $P_n(\eta(x);\bm{\lambda})$ etc.
We consider the {\em shape-invariant\/} \cite{genden} original systems
only, which are characterised by the condition:
\begin{equation}
  \mathcal{A}(\bm{\lambda})\mathcal{A}(\bm{\lambda})^{\dagger}
  =\mathcal{A}(\bm{\lambda}+\bm{\delta})^{\dagger}
  \mathcal{A}(\bm{\lambda}+\bm{\delta})
  +\mathcal{E}_1(\bm{\lambda}),
  \label{shape1}
\end{equation}
or equivalently
\begin{equation}
  \bigl(\partial_xw(x;\bm{\lambda})\bigr)^2
  -\partial_x^2w(x;\bm{\lambda})
  =\bigl(\partial_xw(x;\bm{\lambda}+\bm{\delta})\bigr)^2
  +\partial_x^2w(x;\bm{\lambda}+\bm{\delta})
  +\mathcal{E}_1(\bm{\lambda}).
    \label{shape2}
\end{equation}
Here $\bm{\delta}$ is the shift of the parameters.

It should be remarked that the pair of Hamiltonians,
$\mathcal{A}^{\dagger}\mathcal{A}$ and $\mathcal{A}\mathcal{A}^{\dagger}$
together with the corresponding Darboux-Crum transformations constitute
the basic ingredients of the supersymmetric quantum mechanics \cite{susyqm}.
The series of multiple Darboux-Crum transformations
\eqref{newschr}--\eqref{inpro} can also be formulated in the language of
supersymmetric quantum mechanics.
We believe the formulation in \S\ref{sec:genstr} is simpler and more
direct than that of susy quantum mechanics.

The explicit extensions depend on the choices of the seed solutions
$\{\varphi_j(x), \tilde{\mathcal{E}}_j\}$ ($j=1,\ldots,M$).
The obvious choices are a  subset of the discrete eigenfunctions
$\{\phi_j(x), \mathcal{E}_j\}$ ($j\in \mathcal{D}$), in which $\mathcal{D}$
is a subset of the index set of the total discrete eigenfunctions.
By using those from the groundstate on $\phi_0$, $\phi_1$,\ldots, $\phi_{M-1}$
($\mathcal{D}=\{0,1,\ldots,M-1\}$) successively, Crum \cite{crum} has
derived an essentially iso-spectral extension
\begin{align}
  &\mathcal{H}^{[M]}\phi_n^{[M]}(x)=\mathcal{E}_n\phi_n^{[M]}(x)\quad
  (n=M,M+1,\ldots),\\
  &\phi_n^{[M]}(x)\eqdef
  \frac{\text{W}[\phi_0,\phi_1,\ldots,\phi_{M-1},\phi_n](x)}
  {\text{W}[\phi_0,\phi_1,\ldots,\phi_{M-1}](x)},\quad
  (\phi_m^{[M]},\phi_n^{[M]})
  =\prod_{j=0}^{M-1}(\mathcal{E}_n-\mathcal{E}_{j})\cdot h_n\delta_{m\,n},\\
  &U^{[M]}(x)\eqdef U(x)-2\partial_x^2\log
  \bigl|\text{W}[\phi_0,\phi_1,\ldots,\phi_{M-1}](x)\bigr|,
\end{align}
in which the potential $U^{[s]}(x)$ ($s=1,2,\ldots,M$) at each step is
non-singular. Shape-invariance means simply
\begin{align}
  U^{[s]}(x;\bm{\lambda})&=U(x;\bm{\lambda}+s\bm{\delta})
  +\mathcal{E}_s(\bm{\lambda})\quad(s=1,2,\ldots,M),
  \label{crumshape}\\
  \phi^{[s]}_{s+n}(x;\bm{\lambda})&\propto
  \phi_n(x;\bm{\lambda}+s\bm{\delta})\quad
  (s=1,2,\ldots,M;\ \ n=0,1,\ldots,).
  \label{crumshape2}
\end{align}
By allowing gaps in the final spectrum, Krein and Adler \cite{adler} have
generalised Crum's results for $\mathcal{D}\eqdef\{d_1,d_2,\ldots,d_M\}$
($d_j\in\mathbb{Z}_{\ge0}$ : mutually distinct):
\begin{align}
  &\mathcal{H}^{[M]}\phi_n^{[M]}(x)=\mathcal{E}_n\phi_n^{[M]}(x)\quad
  (n\notin\mathcal{D}),
  \label{KAeq}\\
  &\phi_n^{[M]}(x)\eqdef
  \frac{\text{W}[\phi_{d_1},\phi_{d_2},\ldots,\phi_{d_M},\phi_n](x)}
  {\text{W}[\phi_{d_1},\phi_{d_2},\ldots,\phi_{d_M}](x)},\quad
  (\phi_m^{[M]},\phi_n^{[M]})
  =\prod_{j=1}^M(\mathcal{E}_n-\mathcal{E}_{d_j})\cdot h_n\delta_{m\,n},
  \label{Mnorm}\\
  &U^{[M]}(x)\eqdef U(x)-2\partial_x^2\log
  \bigl|\text{W}[\phi_{d_1},\phi_{d_2},\ldots,\phi_{d_M}](x)\bigr|.
  \label{KAU}
\end{align}
The potential $U^{[M]}(x)$ is non-singular when the set $\mathcal{D}$,
which specifies the gaps, $\phi_{d_j}^{[M]}(x)\equiv0$ ($d_j\in\mathcal{D}$),
satisfies the conditions \cite{adler,gos}:
\begin{equation}
  \text{Krein-Adler conditions :}\quad
  \prod_{j=1}^M(m-d_j)\ge0\quad(\,\forall m\in\mathbb{Z}_{\ge 0}).
  \label{KAcond}
\end{equation}
These simply mean that the gaps are even numbers of consecutive levels.
These extensions are well-known.
In the next subsection, we consider seed functions which are not eigenfunctions.

\subsection{Virtual \& pseudo virtual states}
\label{sec:vir}

Now let us consider extensions in terms of {\em non-eigen seed functions}.
For simplicity sake, we first consider the cases of {\em exactly iso-spectral
extensions} ({\em deformations}).
The seed functions $\{\varphi_j(x),\tilde{\mathcal{E}}_j\}$ ($j=1,2,\ldots,M$)
satisfying the following conditions are called
{\em virtual state wavefunctions\/}
\footnote{
They are different from the `virtual energy levels' in quantum scattering
theory \cite{schiff}.}:
\begin{enumerate}
\item No zeros in $x_1<x<x_2$, {\em i.e.} $\varphi_j(x)>0$ or
$\varphi_j(x)<0$ in $x_1<x<x_2$.
\item Negative energy, $\tilde{\mathcal{E}}_j<0$.
\item $\varphi_j(x)$ is also a polynomial type solution,
like the original eigenfunctions, see \eqref{grab}.
\item Square non-integrability, $(\varphi_j,\varphi_j)=\infty$.
\item Reciprocal square non-integrability,
$(\varphi_j^{-1},\varphi_j^{-1})=\infty$.
\end{enumerate}
Of course these conditions are not totally independent.
The negative energy condition is necessary for the positivity of the norm
as seen from the norm formula \eqref{Mnorm}, since a similar formula is
valid for the virtual state wavefunction cases when $\mathcal{E}_{d_j}$
is replaced by $\tilde{\mathcal{E}}_j$.

When the first condition is dropped and the reciprocal is required to be
square integrable at both boundaries, $(x_1,x_1+\epsilon)$,
$(x_2-\epsilon,x_2)$, $\epsilon>0$, see \eqref{type3},
such seed functions are called {\em pseudo virtual state wavefunctions}.
When the system is extended in terms of a pseudo virtual state wavefunction
$\varphi_j(x)$, the new Hamiltonian $\mathcal{H}^{[1]}$  has an extra
{\em eigenstate} $\varphi_j^{-1}(x)$ with the eigenvalue
$\tilde{\mathcal{E}}_j$, {\em if the new potential is non-singular}.
The extra state is below the original groundstate and $\mathcal{H}^{[1]}$
is no longer iso-spectral with $\mathcal{H}$. This is a consequence of
\eqref{newschr}. Its non-singularity is not guaranteed, either.
When extended in terms of $M$ pseudo virtual state wavefunctions
$\{\varphi_j(x),\tilde{\mathcal{E}}_j\}$ ($j=1,2,\ldots,M$), the resulting
Hamiltonian $\mathcal{H}^{[M]}$ has $M$ additional {\em eigenstates}
$\breve{\varphi}_j^{[M]}(x)$ \eqref{varphiM}, {\em if the potential
$U^{[M]}(x)$ is non-singular}. They are all below the original groundstate.

Since $\varphi_j(x)$ is finite in $x_1<x<x_2$, the non-square integrability
can only be caused by the boundaries. Thus the virtual state wavefunctions
belong to either of the following type $\I$ and $\II$ and the pseudo virtual
state wavefunctions belong to type $\III$ :
\begin{alignat}{3}
  \text{Type $\I$}:&&
  &\int_{x_1}^{x_1+\epsilon}\!\!\!dx\,\varphi_j(x)^2<\infty,\quad
  &&\int_{x_2-\epsilon}^{x_2}\!\!\!dx\,\varphi_j(x)^2=\infty,\n
  &\ \ \text{\&}\ &&\int_{x_1}^{x_1+\epsilon}\!\!\!dx\,\varphi_j(x)^{-2}=\infty,
  \quad
  &&\int_{x_2-\epsilon}^{x_2}\!\!\!dx\,\varphi_j(x)^{-2}<\infty,
  \label{type1}\\
  \text{Type $\II$}:&&
  &\int_{x_1}^{x_1+\epsilon}\!\!\!dx\,\varphi_j(x)^2=\infty,\quad
  &&\int_{x_2-\epsilon}^{x_2}\!\!\!dx\,\varphi_j(x)^2<\infty,\n
  &\ \ \text{\&}\ &&\int_{x_1}^{x_1+\epsilon}\!\!\!dx\,\varphi_j(x)^{-2}<\infty,
  \quad
  &&\int_{x_2-\epsilon}^{x_2}\!\!\!dx\,\varphi_j(x)^{-2}=\infty,
  \label{type2}\\
  \text{Type $\III$}:&&
  &\int_{x_1}^{x_1+\epsilon}\!\!\!dx\,\varphi_j(x)^2=\infty\ \ \text{or}
  &&\int_{x_2-\epsilon}^{x_2}\!\!\!dx\,\varphi_j(x)^2=\infty,\n
  &\ \ \text{\&}\ &&\int_{x_1}^{x_1+\epsilon}\!\!\!dx\,\varphi_j(x)^{-2}<\infty,
  \quad
  &&\int_{x_2-\epsilon}^{x_2}\!\!\!dx\,\varphi_j(x)^{-2}<\infty.
  \label{type3}
\end{alignat}
An appropriate modification is needed when $x_2=+\infty$ and/or $x_1=-\infty$.
Hereafter we denote the pseudo state wavefunctions by
$\{\tilde{\phi}_\text{v}\}$, which are the main ingredients of this paper.

The Darboux-Crum transformations in terms of type $\I$ and $\II$ virtual state
wavefunctions have been applied to achieve exactly iso-spectral deformations of 
the radial oscillator potential and the Darboux-P\"oschl-Teller potentials
\cite{os25}, which  generate the multi-indexed Laguerre and Jacobi polynomials.

In this paper we show that the Darboux-Crum transformations in terms of
a set of $\mathcal{D}\eqdef\{d_1,d_2,\ldots,d_M\}$ pseudo virtual state
wavefunctions
\begin{align}
  &\mathcal{H}^{[M]}\phi_n^{[M]}(x)=\mathcal{E}_n\phi_n^{[M]}(x),
  \label{vDarb}\\
  &\phi_n^{[M]}(x)\eqdef
  \frac{\text{W}[\tilde{\phi}_{d_1},\tilde{\phi}_{d_2},\ldots,
  \tilde{\phi}_{d_M},\phi_n](x)}
  {\text{W}[\tilde{\phi}_{d_1},\tilde{\phi}_{d_2},\ldots,
  \tilde{\phi}_{d_M}](x)},\quad
  (\phi_m^{[M]},\phi_n^{[M]})
  =\prod_{j=1}^M(\mathcal{E}_n-\tilde{\mathcal{E}}_{d_j})\cdot
  h_n\delta_{m\,n},
  \label{Mnorm2}\\
  &U^{[M]}(x)\eqdef U(x)-2\partial_x^2\log
  \bigl|\text{W}[\tilde{\phi}_{d_1},\tilde{\phi}_{d_2},\ldots,
  \tilde{\phi}_{d_M}](x)\bigr|,
  \label{vpot}
\end{align}
are equivalent to a system generated by Krein-Adler transformations
\eqref{KAeq}--\eqref{KAU} specified by a certain set $\bar{\mathcal{D}}$
\eqref{barD} of eigenfunctions with shifted parameters for various systems with
shape-invariant potentials, which are listed in the subsequent section.
In contrast to the extensions in terms of the virtual states, these extensions 
in terms of pseudo virtual states are not {\em iso-spectral} and the obtained
systems are not {\em shape-invariant\/}.

As seen in each example, the virtual and pseudo virtual wavefunctions are
generated from the eigenstate wavefunctions through twisting of parameters
based on the  {\em discrete symmetries} of the original Hamiltonian.

\section{Examples of Shape-invariant Quantum Mechanical Systems}
\label{sec:Exa}

Here we provide the essence of shape-invariant and thus exactly solvable
one-dimensional quantum mechanical systems. The first five examples
\S\,\ref{sec:H}--\S\,\ref{sec:K} have infinitely many discrete eigenstates,
whereas the rest \S\,\ref{sec:M}--\S\,\ref{sec:hDPT} has finitely many
eigenstates.
They are divided into two groups of eigenfunction patterns.
The eigenfunctions $\{\phi_n\}$ and the pseudo virtual state wavefunctions
$\{\tilde{\phi}_\text{v}\}$
\begin{align}
  \mathcal{H}(\bm{\lambda})\phi_n(x;\bm{\lambda})
  &=\mathcal{E}_n(\bm{\lambda})\phi_n(x;\bm{\lambda})
  \quad(\,0\le n\le n_{\text{max}}(\bm{\lambda})\ \ \text{or}
  \ \ n\in\mathbb{Z}_{\ge0}),
  \label{orisys}\\
  \mathcal{H}(\bm{\lambda})\tilde{\phi}_{\text{v}}(x;\bm{\lambda})
  &=\tilde{\mathcal{E}}_{\text{v}}(\bm{\lambda})
  \tilde{\phi}_{\text{v}}(x;\bm{\lambda})
  \quad\bigl(\text{v}\in\mathcal{V}(\bm{\lambda})\bigr),
\end{align}
are listed in detail for reference purposes.
Here $\mathcal{V}(\bm{\lambda})$ is the index set of the pseudo virtual
state wavefunctions,
which is specified for each example. The type $\I$ and $\II$ virtual
state wavefunctions for systems with finitely many eigenstates will be
discussed in a separate paper \cite{os28}.

The basic tools of shape-invariant systems \eqref{shape1}--\eqref{shape2}
are the forward shift and backward shift relations:
\begin{align}
  \mathcal{A}(\bm{\lambda})\phi_n(x;\bm{\lambda})
  &=f_n(\bm{\lambda})\phi_{n-1}(x;\bm{\lambda}+\bm{\delta}),
  \label{forward}\\
  \mathcal{A}(\bm{\lambda})^{\dagger}\phi_{n-1}(x;\bm{\lambda}+\bm{\delta})
  &=b_{n-1}(\bm{\lambda})\phi_n(x;\bm{\lambda}),
  \label{backward}\\
  \mathcal{E}_n(\bm{\lambda})&=f_n(\bm{\lambda})b_{n-1}(\bm{\lambda}).
  \label{enegyfact}
\end{align}
For the parameters of shape-invariant transformation
$\bm{\lambda}\to\bm{\lambda}+\bm{\delta}$, we use the symbol $g$ to denote
an increasing ($\bm{\delta}=1$) parameter and the symbol $h$ for a
decreasing ($\bm{\delta}=-1$) parameter and $\mu$ for an unchanging
($\bm{\delta}=0$) parameter, except for the  Darboux-P\"{o}schl-Teller
potentials (J) in which $g$ and $h$ are both increasing parameters.
Throughout this paper we assume that the parameters $g$ and/or $h$ take
{\em generic values\/}, that is not integers or half odd integers.

In the next subsection, we introduce two groups of the eigenfunction
patterns of the eleven examples of shape-invariant systems,
\S\,\ref{sec:H}--\S\,\ref{sec:hDPT}. Then the basics of the Wronskians
for the two groups are discussed in \S\,\ref{sec:Wro}.
The explicit formulas of the eigenvalues, eigenfunctions,
pseudo virtual state wavefunctions and the discrete symmetries of the
eleven shape-invariant systems are provided in
\S\,\ref{sec:H}--\S\,\ref{sec:hDPT} for reference purposes.
Among them, we also report type $\II$ virtual state wavefunctions for two
potentials (C) \S\,\ref{sec:C} and (Kh) \S\,\ref{sec:Kh}.
These have been reported in \cite{grandati} and \cite{quesne5}, respectively
in connection with the $M=1$ rational extensions.
We stress that multi-indexed orthogonal polynomials can be constructed for
these two potentials in exactly the same way as in \cite{os25}.
For another type of virtual state wavefunctions for the potentials with
finitely many discrete eigenstates \S\,\ref{sec:M}--\S\,\ref{sec:hDPT},
see a subsequent publication \cite{os28}.
The sections \ref{sec:H}--\ref{sec:hDPT} could be skipped in the first reading. 

\subsection{Two groups of eigenfunctions patterns}
\label{sec:2group}

In all the eleven Hamiltonian systems \S\,\ref{sec:H}--\S\,\ref{sec:hDPT}
the energy formula of the pseudo virtual state is related to that of
the `eigenstate' with a negative `degree':
\begin{equation}
  \tilde{\mathcal{E}}_{\text{v}}(\bm{\lambda})
  =\mathcal{E}_{-\text{v}-1}(\bm{\lambda}).
\end{equation}
These Hamiltonian systems are divided into two groups of eigenfunction
patterns:
\begin{equation}
  \phi_n(x;\bm{\lambda})=\phi_{0(n)}(x;\bm{\lambda})
  P_n\bigl(\eta(x);\bm{\lambda}\bigr),\quad
  \phi_{0(n)}(x;\bm{\lambda})=\left\{
  \begin{array}{ll}
  \phi_0(x;\bm{\lambda})&:\text{Group A}\\[1pt]
  \phi_0(x;\bm{\lambda}+n\bm{\delta})&:\text{Group B}
  \end{array}\right..
  \label{grab}
\end{equation}
Seven potentials (H), (L), (J), (M), (s), (hst) and (hDPT) belong to group
A and four potentials (C), (K), (RM) and (Kh) belong to group B.
The group A has a simple structure. The `$n$' dependence is carried only
by the degree of the polynomials (except for (M)).
They all satisfy the closure relations \cite{os7} and $\eta(x)$ is called
a sinusoidal coordinate.
The groundstate wavefunctions of this group satisfy
\begin{equation}
  \frac{\phi_0(x;\bm{\lambda}+\bm{\delta})}{\phi_0(x;\bm{\lambda})}
  =\cF^{-1}\frac{d\eta(x)}{dx},\quad
  \cF=\left\{
  \begin{array}{ll}
  1&:\text{H,\,s,\,hst}\\
  2&:\text{L}\\
  -4&:\text{J}\\
  -1&:\text{M}\\
  4&:\text{hDPT}
  \end{array}\right..
  \label{cF}
\end{equation}
The group B has a more complicated structure. The `$n$' dependence appears
also in the other factor and in the parameter of the polynomials $\alpha_n$,
$\beta_n$.

The pseudo virtual state wavefunctions are obtained from the eigenstate
wavefunctions by twisting $\bm{\lambda}\to\mathfrak{t}(\bm{\lambda})$
(and $x\to ix$ for (H) and (L)):
\begin{align}
  &\tilde{\phi}_{\text{v}}(x;\bm{\lambda})
  =\phi_{\text{v}}\bigl(x;\mathfrak{t}(\bm{\lambda})\bigr)
  =\tilde{\phi}_{0(\text{v})}(x;\bm{\lambda})
  \xi_{\text{v}}\bigl(\eta(x);\bm{\lambda}\bigr),\n
  &\tilde{\phi}_{0(\text{v})}(x;\bm{\lambda})=\left\{
  \begin{array}{ll}
  \phi_0\bigl(x;\mathfrak{t}(\bm{\lambda})\bigr)
  &:\text{Group A}\\[1pt]
  \phi_0\bigl(x;\mathfrak{t}(\bm{\lambda})+\text{v}\bm{\delta}\bigr)
  &:\text{Group B}
  \end{array}\right.,\quad
  \xi_{\text{v}}(\eta;\bm{\lambda})
  =P_{\text{v}}\bigl(\eta;\mathfrak{t}(\bm{\lambda})\bigr),
  \label{tphi}
\end{align}
with a slight modification for (H) and (L):
\begin{align}
  \text{(H)}:\ \ &
  \tilde{\phi}_{\text{v}}(x;\bm{\lambda})
  =i^{-\text{v}}\phi_{\text{v}}(ix)
  =\tilde{\phi}_{0(\text{v})}(x)\xi_{\text{v}}\bigl(\eta(x)\bigr),\n
  &\tilde{\phi}_{0(\text{v})}(x)=\phi_0(ix),\quad
  \xi_{\text{v}}(\eta)=i^{-\text{v}}P_{\text{v}}(i\eta),
  \label{tphiH}\\
  \text{(L)}:\ \ &
  \tilde{\phi}_{\text{v}}(x;\bm{\lambda})
  =i^{g-1}\phi_{\text{v}}\bigl(ix;\mathfrak{t}(\bm{\lambda})\bigr)
  =\tilde{\phi}_{0(\text{v})}(x;\bm{\lambda})
  \xi_{\text{v}}\bigl(\eta(x);\bm{\lambda}\bigr),\n
  &\tilde{\phi}_{0(\text{v})}(x;\bm{\lambda})
  =i^{g-1}\phi_0\bigl(ix;\mathfrak{t}(\bm{\lambda})\bigr),\quad
  \xi_{\text{v}}(\eta;\bm{\lambda})
  =P_{\text{v}}\bigl(-\eta;\mathfrak{t}(\bm{\lambda})\bigr).
  \label{tphiL}
\end{align}
The twist operation $\mathfrak{t}$ satisfies
$\mathfrak{t}(\bm{\lambda}+\alpha\bm{\delta})
=\mathfrak{t}(\bm{\lambda})-\alpha\bm{\delta}$ ($\alpha\in\mathbb{C}$).

Corresponding to the forward and backward shift relations of the
eigenfunctions \eqref{forward}--\eqref{backward} and the energy
factorisation formula \eqref{enegyfact}, those for the pseudo virtual
state wavefunction read:
\begin{align}
  &\mathcal{A}(\bm{\lambda})\tilde{\phi}_{\text{v}}(x;\bm{\lambda})
  =-\epsilon\,b_{\text{v}}(-\bm{\lambda})
  \tilde{\phi}_{\text{v}+1}(x;\bm{\lambda}+\bm{\delta}),\quad
  \epsilon=\left\{
  \begin{array}{ll}
  -1&:\text{H}\\
  1&:\text{others}
  \end{array}\right.,
  \label{twistfor}\\
  &\mathcal{A}(\bm{\lambda})^{\dagger}
  \tilde{\phi}_{\text{v}+1}(x;\bm{\lambda}+\bm{\delta})
  =-\epsilon'f_{\text{v}+1}(-\bm{\lambda})
  \tilde{\phi}_{\text{v}}(x;\bm{\lambda}),\quad
  \epsilon'=\left\{
  \begin{array}{ll}
  -1&:\text{L}\\
  1&:\text{others}
  \end{array}\right.,
  \label{twistback}\\
  &\tilde{\mathcal{E}}_{\text{v}}(\bm{\lambda})
  =\epsilon\epsilon'f_{\text{v}+1}(-\bm{\lambda})b_{\text{v}}(-\bm{\lambda})
  =f_{-\text{v}-1}(\bm{\lambda})b_{-\text{v}-2}(\bm{\lambda})
  =\mathcal{E}_{-\text{v}-1}(\bm{\lambda}).
\end{align}

\subsection{Wronskian formulas}
\label{sec:Wro}

Based on the eigenfunction patterns, the Wronskians of the eigenfunctions
and the pseudo virtual state wavefunctions for the set
$\mathcal{D}=\{d_1,d_2,\ldots,d_M\}$
\begin{equation*}
  \text{W}[\phi_{d_1},\phi_{d_2},\ldots,\phi_{d_M}](x;\bm{\lambda}),\quad
  \text{W}[\tilde{\phi}_{d_1},\tilde{\phi}_{d_2},\ldots,
  \tilde{\phi}_{d_M}](x;\bm{\lambda})
\end{equation*}
are reduced to determinant formulas of the polynomials.
Since the latter is obtained from the former by twisting, we present
derivations of the former.

For Group A, the reduction to the Wronskian of the polynomials is achieved
by the Wronskian formulas \eqref{Wformula1} and \eqref{Wformula3}
($\check{P}_n(x;\bm{\lambda})=P_n\bigl(\eta(x);\bm{\lambda}\bigr)$) :
\begin{align}
  &\text{W}[\phi_{d_1},\phi_{d_2},\ldots,\phi_{d_M}](x;\bm{\lambda})
  =\phi_0(x;\bm{\lambda})^M
  \text{W}[\check{P}_{d_1},\check{P}_{d_2},\ldots,\check{P}_{d_M}]
  (x;\bm{\lambda})\n
  &=\phi_0(x;\bm{\lambda})^M
  \Bigl(\frac{d\eta(x)}{dx}\Bigr)^{\frac12M(M-1)}
  \text{W}[P_{d_1},P_{d_2},\ldots,P_{d_M}]\bigl(\eta(x);\bm{\lambda}\bigr).
\end{align}

For Group B, the derivative operator $\partial_x^{j-1}$ in the Wronskian is
replaced by `covariant derivatives'
$\mathcal{A}\bigl(\bm{\lambda}+(j-2)\bm{\delta}\bigr)\cdots
\mathcal{A}\bigl(\bm{\lambda}+\bm{\delta}\bigr)\mathcal{A}(\bm{\lambda})$
\eqref{covW} and use is made of the forward shift relation \eqref{forward}
to obtain
\begin{align}
  &\mathcal{A}\bigl(\bm{\lambda}+(j-2)\bm{\delta}\bigr)\cdots
  \mathcal{A}\bigl(\bm{\lambda}+\bm{\delta}\bigr)
  \mathcal{A}(\bm{\lambda})\phi_n(x;\bm{\lambda})
  =\prod_{i=0}^{j-2}f_{n-i}(\bm{\lambda}+i\bm{\delta})\cdot
  \phi_{n-j+1}\bigl(x;\bm{\lambda}+(j-1)\bm{\delta}\bigr)\n
  &=\prod_{i=0}^{j-2}f_{n-i}(\bm{\lambda}+i\bm{\delta})\cdot
  \phi_0(x;\bm{\lambda}+n\bm{\delta})
  P_{n-j+1}\bigl(\eta(x);\bm{\lambda}+(j-1)\bm{\delta}\bigr).
\end{align}
We have
\begin{align}
  &\text{W}[\phi_{d_1},\phi_{d_2},\ldots,\phi_{d_M}](x;\bm{\lambda})
  =\det\Bigl(\prod_{i=0}^{j-2}f_{d_k-i}(\bm{\lambda}+i\bm{\delta})\cdot
  \phi_{d_k-j+1}\bigl(x;\bm{\lambda}+(j-1)\bm{\delta}\bigr)
  \Bigr)_{1\leq j,k\leq M}\n
  &=\prod_{k=1}^M\phi_0(x;\bm{\lambda}+d_k\bm{\delta})\cdot
  \det\Bigl(\prod_{i=0}^{j-2}f_{d_k-i}(\bm{\lambda}+i\bm{\delta})\cdot
  P_{d_k-j+1}\bigl(\eta(x);\bm{\lambda}+(j-1)\bm{\delta}\bigr)
  \Bigr)_{1\leq j,k\leq M}.
\end{align}
This method is applicable to Group A, too, in which
$\prod_{k=1}^M\phi_0(x;\bm{\lambda}+d_k\bm{\delta})$ is replaced by
\begin{equation*}
  \prod_{j=1}^M\phi_0\bigl(x;\bm{\lambda}+(j-1)\bm{\lambda}\bigr)
  =\phi_0(x;\bm{\lambda})^M
  \Bigl(\cF^{-1}\frac{d\eta(x)}{dx}\Bigr)^{\frac12M(M-1)}.
\end{equation*}

Let us summarise the results:
\begin{align}
  &\text{W}[\phi_{d_1},\phi_{d_2},\ldots,\phi_{d_M}](x;\bm{\lambda})
  =\bar{A}_{\mathcal{D}}(x;\bm{\lambda})
  \bar{\Xi}_{\mathcal{D}}\bigl(\eta(x);\bm{\lambda}\bigr),
  \label{Wphi=AXi}\\
  &\bar{A}_{\mathcal{D}}(x;\bm{\lambda})\eqdef\left\{
  \begin{array}{ll}
  \phi_0(x;\bm{\lambda})^M
  \bigl(\cF^{-1}\frac{d\eta(x)}{dx}\bigr)^{\frac12M(M-1)}
  &:\text{Group A}\\[4pt]
  \prod_{k=1}^M\phi_0(x;\bm{\lambda}+d_k\bm{\delta})
  &:\text{Group B}
  \end{array}\right.,\\
  &\bar{\Xi}_{\mathcal{D}}(\eta;\bm{\lambda})\eqdef\left\{
  \begin{array}{ll}
  \cF^{\,\frac12M(M-1)}
  \text{W}[P_{d_1},P_{d_2},\ldots,P_{d_M}](\eta;\bm{\lambda})
  &:\text{Group A}\\[4pt]
  \det\bigl(\bar{X}^{\mathcal{D}}_{j,k}(\eta;\bm{\lambda})
  \bigr)_{1\leq j,k\leq M}
  &:\text{Group B}
  \end{array}\right.,\\
  &\bar{X}^{\mathcal{D}}_{j,k}(\eta;\bm{\lambda})\eqdef
  \prod_{i=0}^{j-2}f_{d_k-i}(\bm{\lambda}+i\bm{\delta})\cdot
  P_{d_k-j+1}\bigl(\eta;\bm{\lambda}+(j-1)\bm{\delta}\bigr).
\end{align}
The pseudo virtual state wavefunction Wronskian is simply obtained by
the twisting:
\begin{align}
  &\text{W}[\tilde{\phi}_{d_1},\tilde{\phi}_{d_2},\ldots,
  \tilde{\phi}_{d_M}](x;\bm{\lambda})
  =A_{\mathcal{D}}(x;\bm{\lambda})
  \Xi_{\mathcal{D}}\bigl(\eta(x);\bm{\lambda}\bigr),
  \label{Wtphi=AXi}\\
  &A_{\mathcal{D}}(x;\bm{\lambda})\eqdef\left\{
  \begin{array}{ll}
  \tilde{\phi}_{0(0)}(x;\bm{\lambda})^M
  \bigl(\cF^{-1}\frac{d\eta(x)}{dx}\bigr)^{\frac12M(M-1)}
  &:\text{Group A}\\[4pt]
  \prod_{k=1}^M\phi_0\bigl(x;\mathfrak{t}(\bm{\lambda})+d_k\bm{\delta}\bigr)
  &:\text{Group B}
  \end{array}\right.,\\
  &\Xi_{\mathcal{D}}(\eta;\bm{\lambda})\eqdef\left\{
  \begin{array}{ll}
  \cF^{\,\frac12M(M-1)}
  \text{W}[\xi_{d_1},\xi_{d_2},\ldots,\xi_{d_M}](\eta;\bm{\lambda})
  &:\text{Group A}\\[4pt]
  \det\bigl(X^{\mathcal{D}}_{j,k}(\eta;\bm{\lambda})\bigr)_{1\leq j,k\leq M}
  &:\text{Group B}
  \end{array}\right.,\\
  &X^{\mathcal{D}}_{j,k}(\eta;\bm{\lambda})\eqdef
  \prod_{i=0}^{j-2}f_{d_k-i}\bigl(\mathfrak{t}(\bm{\lambda})
  +i\bm{\delta}\bigr)\cdot
  \xi_{d_k-j+1}\bigl(\eta;\bm{\lambda}-(j-1)\bm{\delta}\bigr).
\end{align}
(Note that the expressions of $\bar{\Xi}_{\mathcal{D}}$ and
$\Xi_{\mathcal{D}}$ for Group B are valid for Group A, too.)
Their relations are
\begin{equation}
  A_{\mathcal{D}}(x;\bm{\lambda})
  =\bar{A}_{\mathcal{D}}\bigl(x;\mathfrak{t}(\bm{\lambda})\bigr),\quad
  \Xi_{\mathcal{D}}(\eta;\bm{\lambda})
  =\bar{\Xi}_{\mathcal{D}}\bigl(\eta;\mathfrak{t}(\bm{\lambda})\bigr),
\end{equation}
with a slight modification for (H) and (L):
\begin{align*}
  \text{H}:\ \ &
  A_{\mathcal{D}}(x;\bm{\lambda})=\bar{A}_{\mathcal{D}}(ix),\quad
  \Xi_{\mathcal{D}}(\eta;\bm{\lambda})
  =i^{-\ell_{\mathcal{D}}}\bar{\Xi}_{\mathcal{D}}(i\eta),\\
  \text{L}:\ \ &
  A_{\mathcal{D}}(x;\bm{\lambda})
  =i^{(g-\frac{M+1}{2})M}
  \bar{A}_{\mathcal{D}}\bigl(ix;\mathfrak{t}(\bm{\lambda})\bigr),\quad
  \Xi_{\mathcal{D}}(\eta;\bm{\lambda})
  =i^{M(M-1)}
  \bar{\Xi}_{\mathcal{D}}\bigl(-\eta;\mathfrak{t}(\bm{\lambda})\bigr).
\end{align*}
Both $\bar{\Xi}_{\mathcal{D}}(\eta;\bm{\lambda})$ and
$\Xi_{\mathcal{D}}(\eta;\bm{\lambda})$ are polynomials
in $\eta$ and their degrees are generically $\ell_{\mathcal{D}}$:
\begin{equation}
  \ell_{\mathcal{D}}\eqdef\sum_{j=1}^Md_j-\frac12M(M-1).
  \label{ellD}
\end{equation}

\noindent
\underline{Remark} :
Strictly speaking, the notation $\mathcal{D}$ of the polynomials
$\bar{\Xi}_{\mathcal{D}}$ and $\Xi_{\mathcal{D}}$ represents an ordered set.
By changing the order of $d_j$'s, $\bar{\Xi}_{\mathcal{D}}$ and
$\Xi_{\mathcal{D}}$ may change sign.
On the other hand the functions $\bar{A}_{\mathcal{D}}$ and
$A_{\mathcal{D}}$ are invariant under the permutations of $d_j$'s.
In order to avoid excessive appearance of $\pm$ signs in the general
formulas involving $\bar{\Xi}_{\mathcal{D}}$ and $\Xi_{\mathcal{D}}$, etc,
we adopt the following convention.
The formulas in \S\,\ref{sec:main} involving the Wronskians and determinants
of various polynomials depending on $\mathcal{D}$, $\bar{\mathcal{D}}$ and
other sets are understood to be true up to a multiplicative constant $\pm1$
coming from the multi-linearity of the determinants.
This does not affect the main Propositions, {\em e.g.}
\eqref{detiden}--\eqref{genwronide} in Proposition \ref{polywron}.

\subsection{Harmonic oscillator (H)}
\label{sec:H}

The well known system of the harmonic oscillator has infinitely many
eigenstates:
\begin{align*}
  &\bm{\lambda}: \text{none},\quad\bm{\delta}: \text{none},\quad
  -\infty<x<\infty,\\
  &w(x;\bm{\lambda})=-\frac12x^2,\quad
  U(x;\bm{\lambda})=x^2-1,\\
  &\mathcal{E}_n(\bm{\lambda})=2n,\quad\eta(x)=x,\quad
  f_n(\bm{\lambda})=2n,\quad b_{n-1}(\bm{\lambda})=1,\\
  &\phi_n(x;\bm{\lambda})
  =\phi_0(x;\bm{\lambda})P_n\bigl(\eta(x);\bm{\lambda}\bigr),\quad
  \phi_0(x;\bm{\lambda})=e^{-\frac12x^2},\quad
  P_n(\eta;\bm{\lambda})=H_n(\eta),\\
  &h_n(\bm{\lambda})=2^nn!\sqrt{\pi}.
\end{align*}
Here $H_n(x)$ is the Hermite polynomial. The system satisfies the
{\em closure relation} introduced in \cite{os7}.
The pseudo virtual state wavefunctions are obtained by the following
discrete symmetry:
\begin{align*}
  &\mathcal{H}(\bm{\lambda})
  =-\mathcal{H}(\bm{\lambda})\bigl|_{x\to ix}
  +\mathcal{E}_{-1}(\bm{\lambda}),\\
  &\tilde{\phi}_{\text{v}}(x;\bm{\lambda})
  =i^{-\text{v}}\phi_{\text{v}}(ix;\bm{\lambda})
  =e^{\frac12x^2}i^{-\text{v}}H_{\text{v}}(ix)
  \quad(\text{v}\in\mathbb{Z}_{\ge0}),\\
  &\tilde{\mathcal{E}}_{\text{v}}(\bm{\lambda})
  =-\mathcal{E}_{\text{v}}(\bm{\lambda}) 
  +\mathcal{E}_{-1}(\bm{\lambda})
  =\mathcal{E}_{-\text{v}-1}(\bm{\lambda}).
\end{align*}

\subsection{Radial oscillator (L)}
\label{sec:L}

The radial oscillator potential has also infinitely many discrete
eigenstates in the specified parameter range:
\begin{align*}
  &\bm{\lambda}=g,\quad\bm{\delta}=1,\quad
  0<x<\infty,\quad g>\frac12,\\
  &w(x;\bm{\lambda})=-\frac12x^2+g\log x,\quad
  U(x;\bm{\lambda})=x^2+\frac{g(g-1)}{x^2}-(1+2g),\\
  &\mathcal{E}_n(\bm{\lambda})=4n,\quad\eta(x)=x^2,\quad
  f_n(\bm{\lambda})=-2,\quad b_{n-1}(\bm{\lambda})=-2n,\\
  &\phi_n(x;\bm{\lambda})
  =\phi_0(x;\bm{\lambda})P_n\bigl(\eta(x);\bm{\lambda}\bigr),\quad
  \phi_0(x;\bm{\lambda})=e^{-\frac12x^2}x^g,\quad
  P_n(\eta;\bm{\lambda})=L_n^{(g-\frac12)}(\eta),\\
  &h_n(\bm{\lambda})=\frac{1}{2\,n!}\,\Gamma(n+g+\tfrac12).
\end{align*}
Here $L_n^{(\alpha)}(\eta)$ is the Laguerre polynomial.
The system satisfies the closure relation \cite{os7}.

There are three types of discrete symmetries:
\begin{alignat*}{2}
  &\text{type $\I$}:&\quad&\mathcal{H}(\bm{\lambda})
  =-\mathcal{H}(\bm{\lambda})\bigl|_{x\to ix}
  -2(1+2g),\\
  &\text{type $\II$}:&\quad&\mathcal{H}(\bm{\lambda})
  =\mathcal{H}\bigl(\mathfrak{t}(\bm{\lambda})\bigr)
  +2(1-2g),\quad
  \mathfrak{t}(\bm{\lambda})=1-g,\\
  &\text{type $\III$}:&\quad&\mathcal{H}(\bm{\lambda})
  =-\mathcal{H}\bigl(\mathfrak{t}(\bm{\lambda})\bigr)\bigl|_{x\to ix}
  +\mathcal{E}_{-1}(\bm{\lambda}),
\end{alignat*}
and the pseudo virtual states are generated by type $\III$:
\begin{align*}
  &\tilde{\phi}_{\text{v}}(x;\bm{\lambda})
  =i^{g-1}\phi_{\text{v}}\bigl(ix;\mathfrak{t}(\bm{\lambda})\bigr)
  =e^{\frac12x^2}x^{1-g}
  P_{\text{v}}\bigl(-\eta(x);\mathfrak{t}(\bm{\lambda})\bigr)
  \quad(\text{v}\in\mathbb{Z}_{\ge0}),\\
  &\tilde{\mathcal{E}}_{\text{v}}(\bm{\lambda})
  =-\mathcal{E}_{\text{v}}\bigl(\mathfrak{t}(\bm{\lambda})\bigr)
  +\mathcal{E}_{-1}(\bm{\lambda})
  =\mathcal{E}_{-\text{v}-1}(\bm{\lambda}).
\end{align*}
The type $\I$ and $\II$ virtual states are obtained by using type
$\I$ and $\II$ discrete symmetries \cite{os25}.

\subsection{Darboux-P\"{o}schl-Teller (DPT) potential (J)}
\label{sec:J}

The DPT potential has also infinitely many discrete eigenstates in
the specified parameter range:
\begin{align*}
  &\bm{\lambda}=(g,h),\quad\bm{\delta}=(1,1),\quad
  0<x<\frac{\pi}{2},\quad g,h>\frac32,\\
  &w(x;\bm{\lambda})=g\log\sin x+h\log\cos x,\quad
  U(x;\bm{\lambda})=\frac{g(g-1)}{\sin^2x}
  +\frac{h(h-1)}{\cos^2 x}-(g+h)^2,\\
  &\mathcal{E}_n(\bm{\lambda})=4n(n+g+h),\quad\eta(x)=\cos 2x,\quad
  f_n(\bm{\lambda})=-2(n+g+h),\quad b_{n-1}(\bm{\lambda})=-2n,\\
  &\phi_n(x;\bm{\lambda})
  =\phi_0(x;\bm{\lambda})P_n\bigl(\eta(x);\bm{\lambda}\bigr),\quad
  \phi_0(x;\bm{\lambda})=(\sin x)^g(\cos x)^h,\quad
  P_n(\eta;\bm{\lambda})=P_n^{(g-\frac12,h-\frac12)}(\eta),\\
  &h_n(\bm{\lambda})=\frac{\Gamma(n+g+\frac12)\Gamma(n+h+\frac12)}
  {2\,n!\,(2n+g+h)\Gamma(n+g+h)}.
\end{align*}
Here $P_n^{(\alpha,\beta)}(\eta)$ is the Jacobi polynomial.
The system satisfies the  closure relation \cite{os7}.

There are three types of discrete symmetries:
\begin{alignat*}{2}
  &\text{type $\I$}:&\quad&\mathcal{H}(\bm{\lambda})
  =\mathcal{H}\bigl(\mathfrak{t}^{\I}(\bm{\lambda})\bigr)
  +(1+2g)(1-2h),\quad
  \mathfrak{t}^{\I}(\bm{\lambda})=(g,1-h),\\
  &\text{type $\II$}:&\quad&\mathcal{H}(\bm{\lambda})
  =\mathcal{H}\bigl(\mathfrak{t}^{\II}(\bm{\lambda})\bigr)
  +(1-2g)(1+2h),\quad
  \mathfrak{t}^{\II}(\bm{\lambda})=(1-g,h),\\
  &\text{type $\III$}:&\quad&\mathcal{H}(\bm{\lambda})
  =\mathcal{H}\bigl(\mathfrak{t}(\bm{\lambda})\bigr)
  +\mathcal{E}_{-1}(\bm{\lambda}),\quad
  \mathfrak{t}=\mathfrak{t}^{\II}\circ\mathfrak{t}^{\I},
  \ \ \mathfrak{t}(\bm{\lambda})=(1-g,1-h),
\end{alignat*}
and the pseudo virtual states are generated by type $\III$:
\begin{align*}
  &\tilde{\phi}_{\text{v}}(x;\bm{\lambda})
  =\phi_{\text{v}}\bigl(x;\mathfrak{t}(\bm{\lambda})\bigr)
  =(\sin x)^{1-g}(\cos x)^{1-h}
  P_{\text{v}}\bigl(\eta(x);\mathfrak{t}(\bm{\lambda})\bigr)
  \quad(0\leq\text{v}<g+h-1),\\
  &\tilde{\mathcal{E}}_{\text{v}}(\bm{\lambda})
  =\mathcal{E}_{\text{v}}\bigl(\mathfrak{t}(\bm{\lambda})\bigr)
  +\mathcal{E}_{-1}(\bm{\lambda})
  =\mathcal{E}_{-\text{v}-1}(\bm{\lambda}).
\end{align*}
The type $\I$ and $\II$ virtual states are obtained by using type
$\I$ and $\II$ discrete symmetries \cite{os25}.
The Hamiltonian $\mathcal{H}$ has also the `left-right' mirror symmetry
$x\to\frac{\pi}{2}-x$, $g\leftrightarrow h$.

\subsection{Coulomb potential plus the centrifugal barrier (C)}
\label{sec:C}

The system has also infinitely many discrete eigenstates in the specified
parameter range:
\begin{align*}
  &\bm{\lambda}=g,\quad\bm{\delta}=1,\quad
  0<x<\infty,\quad g>\frac12,\\
  &w(x;\bm{\lambda})=g\log x-\frac{x}{g},\quad
  U(x;\bm{\lambda})=\frac{g(g-1)}{x^2}-\frac{2}{x}+\frac{1}{g^2},\\
  &\mathcal{E}_n(\bm{\lambda})=\frac{1}{g^2}-\frac{1}{(g+n)^2},\quad
  \eta(x)=x^{-1},\quad
  f_n(\bm{\lambda})=\frac{-2}{g(g+n)^2},\quad
  b_{n-1}(\bm{\lambda})=\frac{-n(2g+n)}{2g},\\
  &\phi_n(x;\bm{\lambda})
  =e^{-\frac{x}{g+n}}x^{g+n}P_n\bigl(\eta(x);\bm{\lambda}\bigr),\quad
  \phi_0(x;\bm{\lambda})=e^{-\frac{x}{g}}x^g,\quad 
  P_n(\eta;\bm{\lambda})
  =\eta^nL_n^{(2g-1)}\bigl(\tfrac{2}{g+n}\eta^{-1}\bigr),\\
  &h_n(\bm{\lambda})=\Bigl(\frac{g+n}{2}\Bigr)^{2g+2}
  \frac{4}{n!}\,\Gamma(2g+n).
\end{align*}

The discrete symmetry and the pseudo virtual state wavefunctions are:
\begin{align*}
  &\mathcal{H}(\bm{\lambda})
  =\mathcal{H}\bigl(\mathfrak{t}(\bm{\lambda})\bigr)
  +\mathcal{E}_{-1}(\bm{\lambda}),\quad
  \mathfrak{t}(\bm{\lambda})=1-g,\\
  &\tilde{\phi}_{\text{v}}(x;\bm{\lambda})
  =\phi_{\text{v}}\bigl(x;\mathfrak{t}(\bm{\lambda})\bigr)
  =e^{\frac{x}{g-\text{v}-1}}x^{1-g+\text{v}}
  P_{\text{v}}\bigl(\eta(x);\mathfrak{t}(\bm{\lambda})\bigr)
  \quad(0\le\text{v}<g-1),\\
  &\tilde{\mathcal{E}}_{\text{v}}(\bm{\lambda})
  =\mathcal{E}_{\text{v}}\bigl(\mathfrak{t}(\bm{\lambda})\bigr)
  +\mathcal{E}_{-1}(\bm{\lambda})
  =\mathcal{E}_{-\text{v}-1}(\bm{\lambda}).
\end{align*}
For $g-1<\text{v}<2g-1$ ($g>\frac32$), the discrete symmetry generates
the type $\II$ virtual states $\tilde{\phi}_{\text{v}}(x;\bm{\lambda})$
\cite{grandati}.

This system can be obtained from the Kepler problem in spherical space
\S\,\ref{sec:K} in a certain limiting procedure.

\subsection{Kepler problem in spherical space (K)}
\label{sec:K}

The system has also infinitely many discrete eigenstates in the specified
parameter range:
\begin{align*}
  &\bm{\lambda}=(g,\mu),\quad\bm{\delta}=(1,0),\quad
  0<x<\pi,\quad g>\frac32,\ \ \mu>0,\\
  &w(x;\bm{\lambda})=g\log\sin x-\frac{\mu}{g}x,\quad
  U(x;\bm{\lambda})=\frac{g(g-1)}{\sin^2x}-2\mu\cot x+\frac{\mu^2}{g^2}-g^2,\\
  &\mathcal{E}_n(\bm{\lambda})
  =(g+n)^2-g^2+\frac{\mu^2}{g^2}-\frac{\mu^2}{(g+n)^2},\quad
  \eta(x)=\cot x,\n
  &f_n(\bm{\lambda})=\frac{g^2(g+n)^2+\mu^2}{g(g+n)^2},\quad
  b_{n-1}(\bm{\lambda})=\frac{n(2g+n)}{g},\\
  &\phi_n(x;\bm{\lambda})
  =e^{-\frac{\mu}{g+n}x}\bigl(\sin x\bigr)^{g+n}
  P_n\bigl(\eta(x);\bm{\lambda}\bigr),\quad
  \phi_0(x;\bm{\lambda})=e^{-\frac{\mu}{g}x}\bigl(\sin x\bigr)^{g},\\
  &P_n(\eta;\bm{\lambda})=i^{-n}P^{(\alpha_n,\beta_n)}_n(i\eta),\quad
  \alpha_n=-g-n+\frac{\mu}{g+n}i,\ \ \beta_n=-g-n-\frac{\mu}{g+n}i,\\
  &h_n(\bm{\lambda})
  =\frac{e^{-\frac{\pi\mu}{g+n}}2^{1-2(g+n)}\pi(g+n)\Gamma(2g+n)}
  {n!\bigl((g+n)^2+\frac{\mu^2}{(g+n)^2}\bigr)
  \Gamma(g+\frac{\mu}{g+n}i)\Gamma(g-\frac{\mu}{g+n}i)}.
\end{align*}

The discrete symmetry and the pseudo virtual state wavefunctions are:
\begin{align*}
  &\mathcal{H}(\bm{\lambda})
  =\mathcal{H}\bigl(\mathfrak{t}(\bm{\lambda})\bigr)
  +\mathcal{E}_{-1}(\bm{\lambda}),\quad
  \mathfrak{t}(\bm{\lambda})=(1-g,\mu),\\
  &\tilde{\phi}_{\text{v}}(x;\bm{\lambda})
  =\phi_{\text{v}}\bigl(x;\mathfrak{t}(\bm{\lambda})\bigr)
  =e^{\frac{\mu}{g-\text{v}-1}x}\bigl(\sin x\bigr)^{1-g+\text{v}}
  P_{\text{v}}\bigl(\eta(x);\mathfrak{t}(\bm{\lambda})\bigr)
  \quad(0\le\text{v}<2g-1),\\
%
  &\tilde{\mathcal{E}}_{\text{v}}(\bm{\lambda})
  =\mathcal{E}_{\text{v}}\bigl(\mathfrak{t}(\bm{\lambda})\bigr)
  +\mathcal{E}_{-1}(\bm{\lambda})
  =\mathcal{E}_{-\text{v}-1}(\bm{\lambda}).
\end{align*}

The Coulomb potential plus the centrifugal barrier \S\,\ref{sec:C} is
obtained by the following limit:
\begin{equation*}
  x=\tfrac{\pi}{L}x^{\text{C}},\ \ \mu=\tfrac{L}{\pi},
  \ \ \lim_{L\to\infty}\bigl(\tfrac{\pi}{L}\bigr)^2\mathcal{H}(\bm{\lambda})
  =\mathcal{H}^{\text{C}}(\bm{\lambda}^{\text{C}}),
\end{equation*}
together with the eigenfunctions.

\subsection{Morse potential (M)}
\label{sec:M}

The system has finitely many discrete eigenstates
$0\le n\le n_\text{max}(\bm{\lambda})=[h]'$ in the specified parameter range
($[a]'$ denotes the greatest integer not exceeding and not equal to $a$):
\begin{align*}
  &\bm{\lambda}=(h,\mu),\quad\bm{\delta}=(-1,0),\quad
  -\infty<x<\infty,\quad h,\mu>0,\\
  &w(x;\bm{\lambda})=hx-\mu e^x,\quad
  U(x;\bm{\lambda})=\mu^2e^{2x}-\mu(2h+1)e^x+h^2,\\
  &\mathcal{E}_n(\bm{\lambda})=h^2-(h-n)^2,\quad\eta(x)=e^{-x},\quad
  f_n(\bm{\lambda})=\frac{n-2h}{2\mu},\quad
  b_{n-1}(\bm{\lambda})=-2n\mu,\\
  &\phi_n(x;\bm{\lambda})
  =\phi_0(x;\bm{\lambda})P_n\bigl(\eta(x);\bm{\lambda}\bigr),\quad
  \phi_0(x;\bm{\lambda})=e^{hx-\mu e^x},\\
  &P_n(\eta;\bm{\lambda})
  =(2\mu\eta^{-1})^{-n}L_n^{(2h-2n)}(2\mu\eta^{-1}),\quad
  h_n(\bm{\lambda})=\frac{\Gamma(2h-n+1)}{(2\mu)^{2h}n!\,2(h-n)}.
\end{align*}
The system satisfies the closure relation \cite{os7}.

The discrete symmetry and the pseudo virtual state wavefunctions are:
\begin{align*}
  &\mathcal{H}(\bm{\lambda})
  =\mathcal{H}\bigl(\mathfrak{t}(\bm{\lambda})\bigr)
  +\mathcal{E}_{-1}(\bm{\lambda}),\quad
  \mathfrak{t}(\bm{\lambda})=(-1-h,-\mu),\\
  &\tilde{\phi}_{\text{v}}(x;\bm{\lambda})
  =\phi_{\text{v}}\bigl(x;\mathfrak{t}(\bm{\lambda})\bigr)
  =e^{-(h+1)x+\mu e^x}
  P_{\text{v}}\bigl(\eta(x);\mathfrak{t}(\bm{\lambda})\bigr)
  \quad(\text{v}\in\mathbb{Z}_{\ge0}),\\
  &\tilde{\mathcal{E}}_{\text{v}}(\bm{\lambda})
  =\mathcal{E}_{\text{v}}\bigl(\mathfrak{t}(\bm{\lambda})\bigr)
  +\mathcal{E}_{-1}(\bm{\lambda})
  =\mathcal{E}_{-\text{v}-1}(\bm{\lambda}).
\end{align*}

This system can be obtained from the hyperbolic Darboux-P\"{o}schl-Teller
potential \S\,\ref{sec:hDPT} in a certain limiting procedure.

\subsection{Soliton potential (s)}
\label{sec:s}

The system has finitely many discrete eigenstates
$0\le n\le n_\text{max}(\bm{\lambda})=[h]'$ in the specified parameter range:
\begin{align*}
  &\bm{\lambda}=h,\quad\bm{\delta}=-1,\quad
  -\infty<x<\infty,\quad h>0,\\
  &w(x;\bm{\lambda})=-h\log\cosh x,\quad
  U(x;\bm{\lambda})=-\frac{h(h+1)}{\cosh^2x}+h^2,\\
  &\mathcal{E}_n(\bm{\lambda})=h^2-(h-n)^2,\quad\eta(x)=\sinh x,\quad
  f_n(\bm{\lambda})=h,\quad b_{n-1}(\bm{\lambda})=\frac{n(2h-n)}{h},\\
  &\phi_n(x;\bm{\lambda})
  =\phi_0(x;\bm{\lambda})P_n\bigl(\eta(x);\bm{\lambda}\bigr),\quad
  \phi_0(x;\bm{\lambda})=(\cosh x)^{-h},\\
  &P_n\bigl(\eta(x);\bm{\lambda}\bigr)=(\cosh x)^nP_n^{(h-n,h-n)}(\tanh x),\quad
  h_n(\bm{\lambda})=\frac{2^{2h-2n}\Gamma(h+1)^2}{n!\,(h-n)\Gamma(2h-n+1)}.
\end{align*}
The system satisfies the closure relation \cite{os7}.
One can rewrite $P_n(\eta;\bm{\lambda})$ as
\begin{equation*}
  P_n(\eta;\bm{\lambda})=
  \frac{(h-[\frac{n-1}{2}])_{[\frac{n+1}{2}]}}{(h-n+\frac12)_{[\frac{n+1}{2}]}}
  i^nP_n^{(-h-\frac12,-h-\frac12)}(i\eta),
\end{equation*}
where $[a]$ denotes the greatest integer not exceeding $a$.

The discrete symmetry and the pseudo virtual state wavefunctions are:
\begin{align*}
  &\mathcal{H}(\bm{\lambda})
  =\mathcal{H}\bigl(\mathfrak{t}(\bm{\lambda})\bigr)
  +\mathcal{E}_{-1}(\bm{\lambda}),\quad
  \mathfrak{t}(\bm{\lambda})=-1-h,\\
  &\tilde{\phi}_{\text{v}}(x;\bm{\lambda})
  =\phi_{\text{v}}\bigl(x;\mathfrak{t}(\bm{\lambda})\bigr)
  =(\cosh x)^{h+1}P_{\text{v}}\bigl(\eta(x);\mathfrak{t}(\bm{\lambda})\bigr)
  \quad(\text{v}\in\mathbb{Z}_{\ge0}),\\
  &\tilde{\mathcal{E}}_{\text{v}}(\bm{\lambda})
  =\mathcal{E}_{\text{v}}\bigl(\mathfrak{t}(\bm{\lambda})\bigr)
  +\mathcal{E}_{-1}(\bm{\lambda})
  =\mathcal{E}_{-\text{v}-1}(\bm{\lambda}).
\end{align*}

This system can be obtained from the Rosen-Morse potential \S\,\ref{sec:RM}
by taking $\mu\to 0$ limit.

\subsection{Rosen-Morse potential (RM)}
\label{sec:RM}

This potential is also called Rosen-Morse $\II$ potential.
The system has finitely many discrete eigenstates
$0\le n\le n_\text{max}(\bm{\lambda})=[h-\sqrt{\mu}\,]'$ in the specified
parameter range:
\begin{align*}
  &\bm{\lambda}=(h,\mu),\quad\bm{\delta}=(-1,0),\quad
  -\infty<x<\infty,\quad h>\sqrt{\mu}>0,\\
  &w(x;\bm{\lambda})=-h\log\cosh x-\frac{\mu}{h}x,\quad
  U(x;\bm{\lambda})=-\frac{h(h+1)}{\cosh^2x}+2\mu\tanh x
  +h^2+\frac{\mu^2}{h^2},\\
  &\mathcal{E}_n(\bm{\lambda})=h^2-(h-n)^2
  +\frac{\mu^2}{h^2}-\frac{\mu^2}{(h-n)^2},\quad\eta(x)=\tanh x,\n
  &f_n(\bm{\lambda})=\frac{h^2(h-n)^2-\mu^2}{h(h-n)^2},\quad
  b_{n-1}(\bm{\lambda})=\frac{n(2h-n)}{h},\\
  &\phi_n(x;\bm{\lambda})=e^{-\frac{\mu}{h-n}x}(\cosh x)^{-h+n}
  P_n\bigl(\eta(x);\bm{\lambda}\bigr),\quad 
  \phi_0(x;\bm{\lambda})=e^{-\frac{\mu}{h}x}(\cosh x)^{-h},\n
  &P_n(\eta;\bm{\lambda})=P_n^{(\alpha_n,\beta_n)}(\eta),\quad
  \alpha_n=h-n+\frac{\mu}{h-n},\ \ \beta_n=h-n-\frac{\mu}{h-n},\\
  &h_n(\bm{\lambda})=\frac{2^{2h-2n}(h-n)
  \Gamma(h+\frac{\mu}{h-n}+1)\Gamma(h-\frac{\mu}{h-n}+1)}
  {n!\,\bigl((h-n)^2-\frac{\mu^2}{(h-n)^2}\bigr)\Gamma(2h-n+1)}.
\end{align*}
By taking the limit $\mu\to 0$, the soliton potential \S\,\ref{sec:s} is
obtained.
Based on the symmetry $x\to-x$, $\mu\to-\mu$, positive $\mu$ is selected.

The discrete symmetry and the pseudo virtual state wavefunctions are:
\begin{align*}
  &\mathcal{H}(\bm{\lambda})
  =\mathcal{H}\bigl(\mathfrak{t}(\bm{\lambda})\bigr)
  +\mathcal{E}_{-1}(\bm{\lambda}),\quad
  \mathfrak{t}(\bm{\lambda})=(-1-h,\mu),\\
  &\tilde{\phi}_{\text{v}}(x;\bm{\lambda})
  =\phi_{\text{v}}\bigl(x;\mathfrak{t}(\bm{\lambda})\bigr)
  =e^{\frac{\mu}{h+1+\text{v}}x}(\cosh x)^{h+1+\text{v}}
  P_{\text{v}}\bigl(\eta(x);\mathfrak{t}(\bm{\lambda})\bigr)
  \quad(\text{v}\in\mathbb{Z}_{\ge0}),\\
  &\tilde{\mathcal{E}}_{\text{v}}(\bm{\lambda})
  =\mathcal{E}_{\text{v}}\bigl(\mathfrak{t}(\bm{\lambda})\bigr)
  +\mathcal{E}_{-1}(\bm{\lambda})
  =\mathcal{E}_{-\text{v}-1}(\bm{\lambda}).
\end{align*}

\subsection{Hyperbolic symmetric top $\II$ (hst)}
\label{sec:hstII}

The system has finitely many discrete eigenstates
$0\le n\le n_\text{max}(\bm{\lambda})=[h]'$ in the specified parameter range:
\begin{align*}
  &\bm{\lambda}=(h,\mu),\quad\bm{\delta}=(-1,0),\quad
  -\infty<x<\infty,\quad h,\mu>0,\\
  &w(x;\bm{\lambda})=-h\log\cosh x-\mu\tan^{-1}\sinh x,\n
  &U(x;\bm{\lambda})=\frac{-h(h+1)+\mu^2+\mu(2h+1)\sinh x}{\cosh^2x}+h^2,\\
  &\mathcal{E}_n(\bm{\lambda})=h^2-(h-n)^2,\quad\eta(x)=\sinh x,\quad
  f_n(\bm{\lambda})=\frac{n-2h}{2},\quad b_{n-1}(\bm{\lambda})=-2n,\\
  &\phi_n(x;\bm{\lambda})
  =\phi_0(x;\bm{\lambda})
  P_n\bigl(\eta(x);\bm{\lambda}\bigr),\quad
  \phi_0(x;\bm{\lambda})=e^{-\mu\tan^{-1}\sinh x}(\cosh x)^{-h},\n
  &P_n(\eta;\bm{\lambda})=i^{-n}P_n^{(\alpha,\beta)}(i\eta),\quad
  \alpha=-h-\tfrac12-i\mu,\ \ \beta=-h-\tfrac12+i\mu,\\
  &h_n(\bm{\lambda})=\frac{\pi\Gamma(2h-n+1)}
  {2^{2h}n!\,(h-n)\Gamma(h-n+\frac12+i\mu)\Gamma(h-n+\frac12-i\mu)}.
\end{align*}
The system satisfies the closure relation \cite{os7}.

The discrete symmetry and the pseudo virtual state wavefunctions are:
\begin{align*}
  &\mathcal{H}(\bm{\lambda})
  =\mathcal{H}\bigl(\mathfrak{t}(\bm{\lambda})\bigr)
  +\mathcal{E}_{-1}(\bm{\lambda}),\quad
  \mathfrak{t}(\bm{\lambda})=(-1-h,-\mu),\\
  &\tilde{\phi}_{\text{v}}(x;\bm{\lambda})
  =\phi_{\text{v}}\bigl(x;\mathfrak{t}(\bm{\lambda})\bigr)
  =e^{\mu\tan^{-1}\sinh x}(\cosh x)^{h+1}
  P_{\text{v}}\bigl(\eta(x);\mathfrak{t}(\bm{\lambda})\bigr)
  \quad(\text{v}\in\mathbb{Z}_{\ge0}),\\
  &\tilde{\mathcal{E}}_{\text{v}}(\bm{\lambda})
  =\mathcal{E}_{\text{v}}\bigl(\mathfrak{t}(\bm{\lambda})\bigr)
  +\mathcal{E}_{-1}(\bm{\lambda})
  =\mathcal{E}_{-\text{v}-1}(\bm{\lambda}).
\end{align*}

\subsection{Kepler problem in hyperbolic space (Kh)}
\label{sec:Kh}

This potential is also called Eckart potential.
It has finitely many discrete eigenstates
$0\le n\le n_\text{max}(\bm{\lambda})=[\sqrt{\mu}-g]'$ in the specified
parameter range:
\begin{align*}
  &\bm{\lambda}=(g,\mu),\quad\bm{\delta}=(1,0),\quad
  0<x<\infty,\quad \sqrt{\mu}>g>\frac12,\\
  &w(x;\bm{\lambda})=g\log\sinh x-\frac{\mu}{g}x,\quad
  U(x;\bm{\lambda})=\frac{g(g-1)}{\sinh^2x}
  -2\mu\coth x+g^2+\frac{\mu^2}{g^2},\\
  &\mathcal{E}_n(\bm{\lambda})=g^2-(g+n)^2
  +\frac{\mu^2}{g^2}-\frac{\mu^2}{(g+n)^2},\quad\eta(x)=\coth x,\n
  &f_n(\bm{\lambda})=\frac{\mu^2-g^2(g+n)^2}{g(g+n)^2},\quad
  b_{n-1}(\bm{\lambda})=\frac{n(2g+n)}{g},\\
  &\phi_n(x;\bm{\lambda})
  =e^{-\frac{\mu}{g+n}x}(\sinh x)^{g+n}P_n\bigl(\eta(x);\bm{\lambda}\bigr),
  \quad 
  \phi_0(x;\bm{\lambda})
  =e^{-\frac{\mu}{g}x}(\sinh x)^{g},\n
  &P_n(\eta;\bm{\lambda})=P_n^{(\alpha_n,\beta_n)}(\eta),\quad
  \alpha_n=-g-n+\frac{\mu}{g+n},\ \ \beta_n=-g-n-\frac{\mu}{g+n},\\
  &h_n(\bm{\lambda})=\frac{(g+n)\Gamma(1-g+\frac{\mu}{g+n})\Gamma(2g+n)}
  {2^{2g+2n}n!\,\bigl(\frac{\mu^2}{(g+n)^2}-(g+n)^2\bigr)
  \Gamma(g+\frac{\mu}{g+n})}.
\end{align*}

The discrete symmetry and the pseudo virtual state wavefunctions are:
\begin{align*}
  &\mathcal{H}(\bm{\lambda})
  =\mathcal{H}\bigl(\mathfrak{t}(\bm{\lambda})\bigr)
  +\mathcal{E}_{-1}(\bm{\lambda}),\quad
  \mathfrak{t}(\bm{\lambda})=(1-g,\mu),\\
  &\tilde{\phi}_{\text{v}}(x;\bm{\lambda})
  =\phi_{\text{v}}\bigl(x;\mathfrak{t}(\bm{\lambda})\bigr)
  =e^{\frac{\mu}{g-\text{v}-1}x}(\sinh x)^{-g+\text{v}+1}
  P_{\text{v}}\bigl(\eta(x);\mathfrak{t}(\bm{\lambda})\bigr),\\
  &\tilde{\mathcal{E}}_{\text{v}}(\bm{\lambda})
  =\mathcal{E}_{\text{v}}\bigl(\mathfrak{t}(\bm{\lambda})\bigr)
  +\mathcal{E}_{-1}(\bm{\lambda})
  =\mathcal{E}_{-\text{v}-1}(\bm{\lambda})
  \quad\bigl(0\le\text{v}<g-1,\ \text{v}>\frac{\mu}{g}+g-1\bigr).
\end{align*}
For $g-1<\text{v}<2g-1$ ($g>\frac32$), the discrete symmetry generates
the type $\II$ virtual states $\tilde{\phi}_{\text{v}}(x;\bm{\lambda})$
\cite{quesne5}.

\subsection{Hyperbolic Darboux-P\"{o}schl-Teller potential (hDPT)}
\label{sec:hDPT}

It has finitely many discrete eigenstates
$0\le n\le n_\text{max}(\bm{\lambda})=[\frac{h-g}{2}]'$ in the specified
parameter range:
\begin{align*}
  &\bm{\lambda}=(g,h),\quad\bm{\delta}=(1,-1),\quad
  0<x<\infty,\quad h>g>\frac12,\\
  &w(x;\bm{\lambda})=g\log\sinh x-h\log\cosh x,\quad
  U(x;\bm{\lambda})=\frac{g(g-1)}{\sinh^2x}
  -\frac{h(h+1)}{\cosh^2 x}+(h-g)^2,\\
  &\mathcal{E}_n(\bm{\lambda})=4n(h-g-n),\quad\eta(x)=\cosh 2x,\quad
  f_n(\bm{\lambda})=2(n+g-h),\quad b_{n-1}(\bm{\lambda})=-2n,\\
  &\phi_n(x;\bm{\lambda})
  =\phi_0(x;\bm{\lambda})P_n\bigl(\eta(x);\bm{\lambda}\bigr),\quad
  \phi_0(x;\bm{\lambda})=(\sinh x)^g(\cosh x)^{-h},\\
  &P_n(\eta;\bm{\lambda})=P_n^{(g-\frac12,-h-\frac12)}(\eta),\quad
  h_n(\bm{\lambda})=\frac{\Gamma(n+g+\frac12)\Gamma(h-g-n+1)}
  {2\,n!\,(h-g-2n)\Gamma(h-n+\frac12)}.
\end{align*}
The eigenvalues can be also expressed as
$\mathcal{E}_n(\bm{\lambda})=4\bigl(\bigl(\tfrac{h-g}{2}\bigr)^2
-\bigl(\tfrac{h-g}{2}-n\bigr)^2\bigr)$.
The system satisfies the closure relation \cite{os7}.

Three types of discrete symmetries are:
\begin{alignat*}{2}
  &\text{type $\I$}:&\quad&\mathcal{H}(\bm{\lambda})
  =\mathcal{H}\bigl(\mathfrak{t}^{\I}(\bm{\lambda})\bigr)
  -(1+2g)(1+2h),\quad
  \mathfrak{t}^{\I}(\bm{\lambda})=(g,-1-h),\\
  &\text{type $\II$}:&\quad&\mathcal{H}(\bm{\lambda})
  =\mathcal{H}\bigl(\mathfrak{t}^{\II}(\bm{\lambda})\bigr)
  -(1-2g)(1-2h),\quad
  \mathfrak{t}^{\II}(\bm{\lambda})=(1-g,h),\\
  &\text{type $\III$}:&\quad&\mathcal{H}(\bm{\lambda})
  =\mathcal{H}\bigl(\mathfrak{t}(\bm{\lambda})\bigr)
  +\mathcal{E}_{-1}(\bm{\lambda}),\quad
  \mathfrak{t}=\mathfrak{t}^{\II}\circ\mathfrak{t}^{\I},
  \ \ \mathfrak{t}(\bm{\lambda})=(1-g,-1-h).
\end{alignat*}
The pseudo virtual state wavefunctions are generated by type $\III$:
\begin{align*}
  &\tilde{\phi}_{\text{v}}(x;\bm{\lambda})
  =\phi_{\text{v}}\bigl(x;\mathfrak{t}(\bm{\lambda})\bigr)
  =(\sinh x)^{1-g}(\cosh x)^{h+1}
  P_{\text{v}}\bigl(\eta(x);\mathfrak{t}(\bm{\lambda})\bigr)
  \quad(\text{v}\in\mathbb{Z}_{\ge0}),\\
  &\tilde{\mathcal{E}}_{\text{v}}(\bm{\lambda})
  =\mathcal{E}_{\text{v}}\bigl(\mathfrak{t}(\bm{\lambda})\bigr)
  +\mathcal{E}_{-1}(\bm{\lambda})
  =\mathcal{E}_{-\text{v}-1}(\bm{\lambda}).
\end{align*}
The type $\I$ and $\II$ virtual states are obtained by using type
$\I$ and $\II$ discrete symmetries.

The Morse potential \S\,\ref{sec:M} is obtained by the following limit:
\begin{equation*}
  x=\tfrac12(x^{\text{M}}-a),
  \ \ g=\tfrac12\mu e^a+\alpha,
  \ \ h=\tfrac12\mu e^a+2h^{\text{M}}+\alpha,\quad
  \lim_{a\to\infty}\mathcal{H}(\bm{\lambda})
  =4\mathcal{H}^{\text{M}}(\bm{\lambda}^{\text{M}}),
\end{equation*}
together with the eigenfunctions.

\section{Main Results}
\label{sec:main}

Let us introduce appropriate symbols and notation for stating the main results.
Let $\mathcal{D}\eqdef\{d_1,d_2,\ldots,d_M\}$ ($d_j\in\mathbb{Z}_{\ge0}$)
be a set of distinct non-negative integers.
We introduce an integer $N$ and fix it to be not less than the maximum of
$\mathcal{D}$:
\begin{equation}
  N\ge\text{max}(\mathcal{D}).
\end{equation}
Let us define a set of distinct non-negative integers
$\bar{\mathcal{D}}=\{0,1,\ldots,N\}\backslash
\{\bar{d}_1,\bar{d}_2,\ldots,\bar{d}_M\}$
together with the shifted parameters $\bar{\bm{\lambda}}$:
\begin{align}
  &\bar{\mathcal{D}}\eqdef\{0,1,\ldots,\breve{\bar{d}}_1,\ldots,
  \breve{\bar{d}}_2,\ldots,\breve{\bar{d}}_M,\ldots,N\}
  =\{e_1,e_2,\ldots,e_{N+1-M}\},\n
  &\bar{d}_j\eqdef N-d_j,\quad
  \bar{\bm{\lambda}}\eqdef \bm{\lambda}-(N+1)\bm{\delta}.
  \label{barD}
\end{align}
Starting from the well-defined original system \eqref{orisys},
the system after Darboux-Crum transformations in terms of a set of pseudo
virtual state wavefunctions $\mathcal{D}$ is described by the Hamiltonian
$\mathcal{H}^{\text{DC}}$
\begin{align}
  &\mathcal{H}^{\text{DC}}=-\frac{d^2}{dx^2}+U^{\text{DC}}(x),\n
  &U^{\text{DC}}(x)=U(x;\bm{\lambda})-2\partial_x^2\log\bigl|
  \text{W}[\tilde{\phi}_{d_1},\tilde{\phi}_{d_2},\ldots,
  \tilde{\phi}_{d_M}](x;\bm{\lambda})\bigr|.
  \label{HDC}
\end{align}
General theory presented in \S\,\ref{sec:darb-crum} states that,
if the Hamiltonian $\mathcal{H}^{\text{DC}}$ is non-singular,
the eigenstates are given by $\Phi^{\text{DC}}_n$ and
$\breve{\Phi}^{\text{DC}}_j$:
\begin{align}
  &\Phi^{\text{DC}}_n(x)=
  \frac{\text{W}[\tilde{\phi}_{d_1},\tilde{\phi}_{d_2},\ldots,
  \tilde{\phi}_{d_M},\phi_n](x;\bm{\lambda})}
  {\text{W}[\tilde{\phi}_{d_1},\tilde{\phi}_{d_2},\ldots,
  \tilde{\phi}_{d_M}](x;\bm{\lambda})}
  \quad(n=0,1,\ldots,n_{\text{max}}(\bm{\lambda})\ \text{or}\ \infty),\n
  &\breve{\Phi}^{\text{DC}}_j(x)=
  \frac{\text{W}[\tilde{\phi}_{d_1},\tilde{\phi}_{d_2},\ldots,
  \breve{\tilde{\phi}}_{d_j},\ldots,\tilde{\phi}_{d_M}](x;\bm{\lambda})}
  {\text{W}[\tilde{\phi}_{d_1},\tilde{\phi}_{d_2},\ldots,
  \tilde{\phi}_{d_M}](x;\bm{\lambda})}
  \quad(j=1,2,\ldots,M),\n
  &\mathcal{H}^{\text{DC}}\Phi^{\text{DC}}_n(x)
  =\mathcal{E}_n(\bm{\lambda})\Phi^{\text{DC}}_n(x),\quad
  \mathcal{H}^{\text{DC}}\breve{\Phi}^{\text{DC}}_j(x)
  =\mathcal{E}_{-d_j-1}(\bm{\lambda})\breve{\Phi}^{\text{DC}}_j(x).
  \label{DCdiffeq}
\end{align}
The system after Krein-Adler transformations in terms of $\bar{\mathcal{D}}$
with shifted parameters $\bar{\bm{\lambda}}$ is described by the Hamiltonian
$\mathcal{H}^{\text{KA}}$
\begin{align}
  &\mathcal{H}^{\text{KA}}=-\frac{d^2}{dx^2}+U^{\text{KA}}(x),\n
  &U^{\text{KA}}(x)=U(x;\bar{\bm{\lambda}})-2\partial_x^2\log\bigl|
  \text{W}[\phi_0,\phi_1,\ldots,\breve{\phi}_{\bar{d}_1},\ldots,
  \breve{\phi}_{\bar{d}_M},\ldots,\phi_N](x;\bar{\bm{\lambda}})\bigr|.
  \label{HKA}
\end{align}
Here we assume that the original system \eqref{orisys} with the shifted
parameters $\bar{\bm{\lambda}}$ has square integrable eigenstates, etc.
If the Krein-Adler conditions are fulfilled, eigenstates are given by
$\Phi^{\text{KA}}_n$ and $\breve{\Phi}^{\text{KA}}_j$:
\begin{align}
  &\Phi^{\text{KA}}_n(x)=
  \frac{\text{W}[\phi_0,\phi_1,\ldots,\breve{\phi}_{\bar{d}_1},
  \ldots,\breve{\phi}_{\bar{d}_M},\ldots,\phi_N,\phi_{N+1+n}]
  (x;\bar{\bm{\lambda}})}
  {\text{W}[\phi_0,\phi_1,\ldots,\breve{\phi}_{\bar{d}_1},\ldots,
  \breve{\phi}_{\bar{d}_M},\ldots,\phi_N](x;\bar{\bm{\lambda}})}
  \quad(n=0,1,\ldots,n_{\text{max}}(\bm{\lambda})\ \text{or}\ \infty),\n
  &\breve{\Phi}^{\text{KA}}_j(x)=
  \frac{\text{W}[\phi_0,\phi_1,\ldots,\breve{\phi}_{\bar{d}_1},
  \ldots,\phi_{\bar{d}_j},\ldots,\breve{\phi}_{\bar{d}_M},\ldots,\phi_N]
  (x;\bar{\bm{\lambda}})}
  {\text{W}[\phi_0,\phi_1,\ldots,\breve{\phi}_{\bar{d}_1},\ldots,
  \breve{\phi}_{\bar{d}_M},\ldots,\phi_N](x;\bar{\bm{\lambda}})}
  \quad(j=1,2,\ldots,M),\n
  &\mathcal{H}^{\text{KA}}\Phi^{\text{KA}}_n(x)
  =\mathcal{E}_{N+1+n}(\bar{\bm{\lambda}})\Phi^{\text{KA}}_n(x),\quad
  \mathcal{H}^{\text{KA}}\breve{\Phi}^{\text{KA}}_j(x)
  =\mathcal{E}_{\bar{d}_j}(\bar{\bm{\lambda}})\breve{\Phi}^{\text{KA}}_j(x).
  \label{KAdiffeq}
\end{align}
Note that $n_{\text{max}}(\bm{\lambda})$ in \S\,\ref{sec:M}--\S\,\ref{sec:hDPT}
satisfies
\begin{equation}
  n_{\text{max}}(\bar{\bm{\lambda}})-(N+1)=n_{\text{max}}(\bm{\lambda}).
\end{equation}
The differential equations $\eqref{DCdiffeq}$ and \eqref{KAdiffeq}
(with $n\in\mathbb{Z}_{\geq 0}$) hold irrespective of the non-singularity
of $\mathcal{H}^{\text{DC}}$ and $\mathcal{H}^{\text{KA}}$.

Our main results hold for any one of the shape-invariant quantum mechanical
systems in \S\,\ref{sec:H}--\S\,\ref{sec:hDPT}.

\begin{prop}\label{main}
The two systems with $\mathcal{H}^{\text{\rm DC}}$ and
$\mathcal{H}^{\text{\rm KA}}$ are equivalent.
To be more specific, the equality of the potentials and the eigenfunctions
read\/{\rm{:}}
\begin{align}
  U^{\text{\rm DC}}(x)-\mathcal{E}_{-N-1}(\bm{\lambda})
  &=U^{\text{\rm KA}}(x),
  \label{poteq}\\
  \Phi^{\text{\rm DC}}_n(x)&\propto\Phi^{\text{\rm KA}}_n(x)
  \quad(n=0,1,\ldots,n_{\text{\rm max}}(\bm{\lambda})\ \text{\rm or}\ \infty),
  \label{eigeq1}\\
  \breve{\Phi}^{\text{\rm DC}}_j(x)&\propto\breve{\Phi}^{\text{\rm KA}}_j(x)
  \quad(j=1,2,\ldots,M).
  \label{eigeq2}
\end{align}
The singularity free conditions of the potential are
\begin{equation}
  \prod_{j=1}^{N+1-M}(m-e_j)\ge0
  \quad(\,\forall m\in\mathbb{Z}_{\geq 0}).
  \label{non-sing}
\end{equation}
For $M=1$, $\mathcal{D}=\{d_1\}$,
$\bar{\mathcal{D}}=\{0,1,\ldots,\breve{\bar{d}}_1,\ldots,N\}$,
the above conditions are satisfied by even $d_1$,
$d_1\in 2\mathbb{Z}_{\geq 0}$.
In other words, the pseudo virtual state wavefunctions
$\{\tilde{\phi}_{\text{\rm v}}\}$ for even $\text{\rm v}$ are nodeless.
The above equalities (up to multiplicative factors)
\eqref{poteq}--\eqref{eigeq2} (\eqref{eigeq1} with $n\in\mathbb{Z}_{\geq 0}$)
are algebraic and they hold irrespective of the non-singularity conditions
\eqref{non-sing}.
\end{prop}
The following energy formulas satisfied by the eleven systems
\begin{align}
  \mathcal{E}_{n}(\bm{\lambda})
  -\mathcal{E}_{-N-1}(\bm{\lambda})
  &=\mathcal{E}_{N+1+n}(\bar{\bm{\lambda}}),\n
  \mathcal{E}_{-\text{v}-1}(\bm{\lambda})
  -\mathcal{E}_{-N-1}(\bm{\lambda})
  &=\mathcal{E}_{N-\text{v}}(\bar{\bm{\lambda}}),
  \label{enformula}
\end{align}
and the illustration (Figure 1) would be helpful to understand the Proposition.
The form of the energy curve is that of the DPT potential but the situation
is similar in all other ten potentials.

\begin{figure}[htbp]
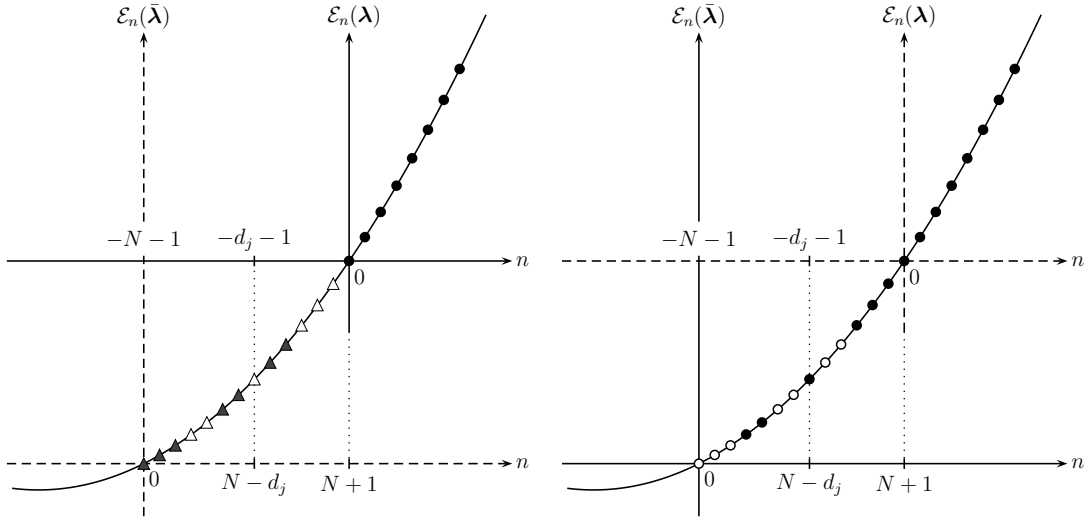

\begin{center}
  \scalebox{0.7}{\includegraphics{enfigIII_pvs.epsi}}\quad
  \scalebox{0.7}{\includegraphics{enfigIII_es.epsi}}
  \caption{The left represents the Darboux-Crum transformations in terms of
pseudo virtual states. The right corresponds to the Krein-Adler
transformations in terms of eigenstates with shifted parameters.
The black circles denote eigenstates.
The white circles in the right graphic denote deleted eigenstates.
The white triangles in the left graphic denote the pseudo virtual states
used in the Darboux-Crum transformations \eqref{DCdiffeq}.
The black triangles denote the unused pseudo virtual states.
}
\end{center}
\end{figure}

If the equality of the potentials \eqref{poteq} is shown, then, under the
condition \eqref{non-sing}, the Hamiltonians
$\mathcal{H}^{\text{DC}}-\mathcal{E}_{-N-1}(\bm{\lambda})
=\mathcal{H}^{\text{KA}}$ are non-singular and it implies \eqref{eigeq1}
and \eqref{eigeq2}.
Therefore the remaining task is to show \eqref{poteq}.

The basic ingredients of the Proposition \ref{main} are the two Wronskians
\begin{equation*}
  \text{W}[\tilde{\phi}_{d_1},\tilde{\phi}_{d_2},\ldots,\tilde{\phi}_{d_M}]
  (x;\bm{\lambda}),\quad
  \text{W}[\phi_0,{\phi}_{1},\ldots,\breve{\phi}_{\bar{d}_1},\ldots,
  \breve{\phi}_{\bar{d}_M},\ldots,\phi_N](x;\bar{\bm{\lambda}}),
\end{equation*}
which determine the potentials \eqref{poteq} and eventually all the
eigenfunctions \eqref{eigeq1}--\eqref{eigeq2}.
By using the Wronskian formulas \eqref{Wtphi=AXi} and \eqref{Wphi=AXi}
they are expressed by the determinants of polynomials:
\begin{align}
  &\text{W}[\tilde{\phi}_{d_1},\tilde{\phi}_{d_2},\ldots,
  \tilde{\phi}_{d_M}](x;\bm{\lambda})
  =A_{\mathcal{D}}(x;\bm{\lambda})
  \Xi_{\mathcal{D}}\bigl(\eta(x);\bm{\lambda}\bigr),
  \label{eigwron1}\\
  &\text{W}[\phi_0,{\phi}_{1},\ldots,\breve{\phi}_{\bar{d}_1},\ldots,
  \breve{\phi}_{\bar{d}_M},\ldots,\phi_N](x;\bar{\bm{\lambda}})
  =\bar{A}_{\bar{\mathcal{D}}}(x;\bar{\bm{\lambda}})
  \bar{\Xi}_{\bar{\mathcal{D}}}\bigl(\eta(x);\bar{\bm{\lambda}}\bigr).
  \label{eigwron2}
\end{align}
These two polynomials $\Xi_{\mathcal{D}}(\eta;\bm{\lambda})$ and
$\bar{\Xi}_{\bar{\mathcal{D}}}(\eta;\bar{\bm{\lambda}})$ have the same degree
$\ell_{\mathcal{D}}=\ell_{\bar{\mathcal{D}}}$ \eqref{ellD}.
An important relation between $A_{\mathcal{D}}(x;\bm{\lambda})$ and
$\bar{A}_{\bar{\mathcal{D}}}(x;\bar{\bm{\lambda}})$ is that its ratio is
independent of $M$ and $d_j$ in each of the eleven potentials
\S\,\ref{sec:H}--\S\,\ref{sec:hDPT}, which is shown by straightforward
calculation.
We set the ratio as $F(x,N,\bm{\lambda})$:
\begin{equation}
  \frac{A_{\mathcal{D}}(x;\bm{\lambda})}
  {\bar{A}_{\bar{\mathcal{D}}}(x;\bar{\bm{\lambda}})}
  =F(x,N,\bm{\lambda}).
  \label{A/A=F}
\end{equation}
{}From this independence and the obvious relation
$A_{\{\,\}}(x;\bm{\lambda})=\bar{A}_{\{\,\}}(x;\bar{\bm{\lambda}})=1$,
the ratio $F$ can be expressed as
\begin{equation}
  F(x,N,\bm{\lambda})=A_{\{0,1,\ldots,N\}}(x;\bm{\lambda})
  =\frac{1}{\bar{A}_{\{0,1,\ldots,N\}}(x;\bar{\bm{\lambda}})}.
  \label{Fform}
\end{equation}
It is easy to show the following identity for any of the eleven systems:
\begin{equation}
  U(x;\bm{\lambda})-2\partial_x^2\log\bigl|F(x,N,\bm{\lambda})\bigr|
  -\mathcal{E}_{-N-1}(\bm{\lambda})
  =U(x;\bar{\bm{\lambda}}).
  \label{UFid}
\end{equation}
By this identity, the equality of the potentials \eqref{poteq} reduces to
\begin{equation}
  \partial_x^2\log\bigl|\Xi_{\mathcal{D}}
  \bigl(\eta(x);\bm{\lambda}\bigr)\bigr|=
  \partial_x^2\log\bigl|\bar{\Xi}_{\bar{\mathcal{D}}}
  \bigl(\eta(x);\bar{\bm{\lambda}}\bigr)\bigr|.
  \label{potequiv}
\end{equation}
Since both polynomials are of the same degree $\ell_{\mathcal{D}}$, this
means that the two polynomials are proportional.
We state this as
\begin{prop}\label{polywron}
Two polynomials characterising $\mathcal{H}^{\text{\rm DC}}$ and
$\mathcal{H}^{\text{\rm KA}}$, $\Xi_{\mathcal{D}}(\eta;\bm{\lambda})$ and
$\bar{\Xi}_{\bar{\mathcal{D}}}(\eta;\bar{\bm{\lambda}})$, are proportional:
\begin{equation}
  \Xi_{\mathcal{D}}(\eta;\bm{\lambda})\propto
  \bar{\Xi}_{\bar{\mathcal{D}}}(\eta;\bar{\bm{\lambda}}).
  \label{detiden}
\end{equation}
\end{prop}
In particular, it means {\em polynomial Wronskian identities}
\begin{equation}
  \text{W}[\xi_{d_1},\xi_{d_2},\ldots,\xi_{d_M}](\eta;\bm{\lambda})
  \propto
  \text{W}[P_0,P_{1},\ldots,\breve{P}_{\bar{d}_1},\ldots,
  \breve{P}_{\bar{d}_M},\ldots,P_N](\eta;\bar{\bm{\lambda}}),
  \label{genwronide}
\end{equation}
for all the systems in Group A.
Recall that $\xi_{\text{v}}(\eta;\bm{\lambda})
=P_{\text{v}}(\eta;\mathfrak{t}(\bm{\lambda}))$ \eqref{tphi}
(with a slight modification for (H) and (L), \eqref{tphiH}--\eqref{tphiL}).
For the simplest case of $M=1$, $N=d_1\equiv\ell$
($\Rightarrow$ $\mathcal{D}=\{\ell\}$,
$\bar{\mathcal{D}}=\{1,2,\ldots,\ell\}$), the Wronskian identities are
reported as (A.22) in \cite{gos} for the three types of the classical
orthogonal polynomials, the Hermite (H), Laguerre (L) and Jacobi (J).
To the best of our knowledge, the general polynomial Wronskian identities
\eqref{genwronide} have not been reported before.
 
The proof of Proposition \ref{polywron} is done by induction in $M$.

\noindent
\underline{first step} :
In the first step we prove \eqref{detiden} for $M=1$,
$N\ge d_1\equiv \text{v}$:
\begin{equation}
  \xi_{\text{v}}(\eta;\bm{\lambda})\propto
  \bar{\Xi}_{\{0,1,\ldots,\breve{\bar{\text{v}}},\ldots,N\}}
  (\eta;\bar{\bm{\lambda}}).
\end{equation}
Recall that the differential equations \eqref{DCdiffeq} and \eqref{KAdiffeq}
hold, and
\begin{align}
  &\tilde{\phi}_{\text{v}}(x;\bm{\lambda})=A_{\{\text{v}\}}(x;\bm{\lambda})
  \xi_{\text{v}}\bigl(\eta(x);\bm{\lambda}\bigr),
  \label{M1eq1}\\
  &\text{W}[\phi_0,\phi_1,\ldots,\ldots,\phi_N](x;\bar{\bm{\lambda}})
  =\bar{A}_{\{0,1,\ldots,N\}}(x;\bar{\bm{\lambda}})
  \bar{\Xi}_{\{0,1,\ldots,N\}}(\eta;\bar{\bm{\lambda}}),\n
  &\text{W}[\phi_0,\phi_1,\ldots,\breve{\phi}_{\bar{\text{v}}},\ldots,\phi_N]
  (x;\bar{\bm{\lambda}})
  =\bar{A}_{\{0,1,\ldots,\breve{\bar{\text{v}}},\ldots,N\}}
  (x;\bar{\bm{\lambda}})
  \bar{\Xi}_{\{0,1,\ldots,\breve{\bar{\text{v}}},\ldots,N\}}
  (\eta;\bar{\bm{\lambda}}).
  \label{M1eq2}
\end{align}
By substituting \eqref{M1eq1} into the differential equation
$\mathcal{H}^{\text{DC}}\breve{\Phi}^{\text{DC}}_1(x)
=\mathcal{E}_{-\text{v}-1}(\bm{\lambda})\breve{\Phi}^{\text{DC}}_1(x)$
\eqref{DCdiffeq}, it becomes
\begin{equation}
  \partial_x^2f
  +2\frac{\partial_x A_{\{\text{v}\}}(x;\bm{\lambda})}
  {A_{\{\text{v}\}}(x;\bm{\lambda})}\,\partial_xf
  +\Bigl(\frac{\partial_x^2A_{\{\text{v}\}}(x;\bm{\lambda})}
  {A_{\{\text{v}\}}(x;\bm{\lambda})}
  -U(x;\bm{\lambda})+\mathcal{E}_{-\text{v}-1}(\bm{\lambda})\Bigr)f=0,
  \label{diffeq1}
\end{equation}
where $f=\xi_{\text{v}}\bigl(\eta(x);\bm{\lambda}\bigr)$.
By substituting \eqref{M1eq2} into the differential equation
$\mathcal{H}^{\text{KA}}\breve{\Phi}^{\text{KA}}_1(x)
=\mathcal{E}_{\bar{\text{v}}}(\bar{\bm{\lambda}})\breve{\Phi}^{\text{KA}}_1(x)$
\eqref{KAdiffeq} and using $\bar{\Xi}_{\{0,1,\ldots,N\}}\bigl(\eta(x);
\bar{\bm{\lambda}}\bigr)=\text{constant}$ (see \eqref{Xi=const}), we obtain
\begin{equation}
  \partial_x^2\bar{f}
  +2\partial_x\log\Bigl|\frac{G_{\text{v}}}{G}\Bigr|\cdot\partial_x\bar{f}
  +\Bigl(\frac{\partial_x^2G_{\text{v}}}{G_{\text{v}}}+\frac{\partial_x^2G}{G}
  -2\frac{\partial_xG_{\text{v}}}{G_{\text{v}}}\frac{\partial_xG}{G}
  -U(x;\bm{\bar{\lambda}})
  +\mathcal{E}_{\bar{\text{v}}}(\bm{\bar{\lambda}})\Bigr)\bar{f}=0,
  \label{diffeq2}
\end{equation}
where $\bar{f}=\bar{\Xi}_{\{0,1,\ldots,\breve{\bar{\text{v}}},\ldots,N\}}
\bigl(\eta(x);\bar{\bm{\lambda}}\bigr)$,
$G_{\text{v}}=\bar{A}_{\{0,1,\ldots,\breve{\bar{\text{v}}},\ldots,N\}}
(x;\bar{\bm{\lambda}})$ and
$G=\bar{A}_{\{0,1,\ldots,N\}}(x;\bar{\bm{\lambda}})$.
By using \eqref{A/A=F} and \eqref{Fform}, we find
$A_{\{\text{v}\}}(x;\bm{\lambda})/G_{\text{v}}=F(x,N,\bm{\lambda})=1/G$.
The above equation \eqref{diffeq2} is rewritten as
\begin{equation}
  \partial_x^2\bar{f}
  +2\frac{\partial_x A_{\{\text{v}\}}(x;\bm{\lambda})}
  {A_{\{\text{v}\}}(x;\bm{\lambda})}\,\partial_x\bar{f}
  +\Bigl(\frac{\partial_x^2A_{\{\text{v}\}}(x;\bm{\lambda})}
  {A_{\{\text{v}\}}(x;\bm{\lambda})}
  -U(x;\bar{\bm{\lambda}})+\mathcal{E}_{\bar{\text{v}}}(\bar{\bm{\lambda}})
  -2\partial_x^2\log F(x,N,\bm{\lambda})\Bigr)\bar{f}=0.
  \label{diffeq2'}
\end{equation}
{}From \eqref{enformula} and \eqref{UFid}, we have
\begin{equation*}
  U(x;\bm{\lambda})-\mathcal{E}_{-\text{v}-1}(\bm{\lambda})
  =U(x;\bar{\bm{\lambda}})-\mathcal{E}_{\bar{\text{v}}}(\bar{\bm{\lambda}})
  +2\partial_x^2\log F(x,N,\bm{\lambda}).
\end{equation*}
Thus $f=\xi_{\text{v}}\bigl(\eta(x);\bm{\lambda}\bigr)$ and
$\bar{f}=\bar{\Xi}_{\{0,1,\ldots,\breve{\bar{\text{v}}},\ldots,N\}}
\bigl(\eta(x);\bar{\bm{\lambda}}\bigr)$ satisfy the same differential
equation. Since both of them are polynomials of degree $\text{v}$ in $\eta$,
they should be proportional for generic parameters \cite{cal-ge}.

\bigskip

\noindent
\underline{second step} :
Assume that \eqref{detiden} holds till $M$ ($M\ge1$), we will show that
it also holds for $M+1$.

Before presenting a general proof, we illustrate the outline by taking
the simple case of Group A \eqref{genwronide}.
We have shown $M=1$ case:
\begin{equation*}
  \xi_{\text{v}}(\eta;\bm{\lambda})\propto
  \text{W}[P_0,\ldots,\breve{P}_{\bar{\text{v}}},\ldots,P_N]
  (\eta;\bar{\bm{\lambda}}).
\end{equation*}
By using the algebraic Wronskian identity \eqref{Wformula2}, we have
\begin{align*}
  \text{W}[\xi_{d_1},\xi_{d_2}](\eta;\bm{\lambda})
  &\propto\text{W}\bigl[
  \text{W}[P_0,\ldots,\breve{P}_{\bar{d}_1},\ldots,P_N],
  \text{W}[P_0,\ldots,\breve{P}_{\bar{d}_2},\ldots,P_N]
  \bigr](\eta;\bar{\bm{\lambda}})\\
  &=\pm
  \text{W}[P_0,\ldots,\breve{P}_{\bar{d}_1},\ldots,\breve{P}_{\bar{d}_2},
  \ldots,P_N](\eta;\bar{\bm{\lambda}})\cdot
  \text{W}[P_0,P_1,\ldots,P_N](\eta;\bar{\bm{\lambda}})\\
  &\propto
  \text{W}[P_0,\ldots,\breve{P}_{\bar{d}_1},\ldots,\breve{P}_{\bar{d}_2},
  \ldots,P_N](\eta;\bar{\bm{\lambda}}),
\end{align*}
where we have used
$\text{W}[P_0,P_1,\ldots,P_N](\eta;\bar{\bm{\lambda}})=\text{constant}$
(see \eqref{Xi=const}).
This is the $M=2$ result.
For $M=3$, we use the $M=2$ results to obtain
\begin{align*}
  &\quad\xi_{d_1}(\eta;\bm{\lambda})\cdot
  \text{W}[\xi_{d_1},\xi_{d_2},\xi_{d_3}](\eta;\bm{\lambda})\\
  &=\text{W}\bigl[
  \text{W}[\xi_{d_1},\xi_{d_2}],\text{W}[\xi_{d_1},\xi_{d_3}]\bigr]
  \bigl(\eta;\bm{\lambda})\\
  &\propto\text{W}\bigl[
  \text{W}[P_0,\ldots,\breve{P}_{\bar{d}_1},\ldots,\breve{P}_{\bar{d}_2},
  \ldots,P_N],
  \text{W}[P_0,\ldots,\breve{P}_{\bar{d}_1},\ldots,\breve{P}_{\bar{d}_3},
  \ldots,P_N]\bigl](\eta;\bar{\bm{\lambda}})\\
  &=\pm
  \text{W}[P_0,\ldots,\breve{P}_{\bar{d}_1},\ldots,\breve{P}_{\bar{d}_2},
  \ldots,\breve{P}_{\bar{d}_3},\ldots,P_N](\eta;\bar{\bm{\lambda}})\cdot
  \text{W}[P_0,\ldots,\breve{P}_{\bar{d}_1},\ldots,P_N]
  (\eta;\bar{\bm{\lambda}})\\
  &\propto
  \text{W}[P_0,\ldots,\breve{P}_{\bar{d}_1},\ldots,\breve{P}_{\bar{d}_2},
  \ldots,\breve{P}_{\bar{d}_3},\ldots,P_N](\eta;\bar{\bm{\lambda}})
  \cdot\xi_{d_1}(\eta;\bm{\lambda}).
\end{align*}
This establishes the identities for $M=3$.
Higher $M$ identities follow in a similar way.

\medskip

Let us present a general proof.
Eq.\eqref{detiden} is equivalent to
\begin{equation*}
  \frac{\text{W}[\tilde{\phi}_{d_1},\tilde{\phi}_{d_2},\ldots,
  \tilde{\phi}_{d_M}](x;\bm{\lambda})}
  {A_{\mathcal{D}}(x;\bm{\lambda})}
  \propto
  \frac{\text{W}[\phi_0,{\phi}_{1},\ldots,\breve{\phi}_{\bar{d}_1},\ldots,
  \breve{\phi}_{\bar{d}_M},\ldots,\phi_N](x;\bar{\bm{\lambda}})}
  {\bar{A}_{\bar{\mathcal{D}}}(x;\bar{\bm{\lambda}})},
\end{equation*}
which implies
\begin{equation*}
  \text{W}[\tilde{\phi}_{d_1},\tilde{\phi}_{d_2},\ldots,
  \tilde{\phi}_{d_M}](x;\bm{\lambda})
  \propto F(x,N,\bm{\lambda})
  \text{W}[\phi_0,{\phi}_{1},\ldots,\breve{\phi}_{\bar{d}_1},\ldots,
  \breve{\phi}_{\bar{d}_M},\ldots,\phi_N](x;\bar{\bm{\lambda}}).
\end{equation*}
Assume that \eqref{detiden} holds till $M$ ($M\ge1$).
By using the Wronskian identity \eqref{Wformula2}, we obtain
\begin{align*}
  &\quad\text{W}[\tilde{\phi}_{d_1},\ldots,\tilde{\phi}_{d_{M-1}}]
  (x;\bm{\lambda})
  \cdot\text{W}[\tilde{\phi}_{d_1},\ldots,\tilde{\phi}_{d_{M-1}},
  \tilde{\phi}_{d_M},\tilde{\phi}_{d_{M+1}}](x;\bm{\lambda})\\
  &=\text{W}\bigl[
  \text{W}[\tilde{\phi}_{d_1},\ldots,\tilde{\phi}_{d_{M-1}},
  \tilde{\phi}_{d_M}],
  \text{W}[\tilde{\phi}_{d_1},\ldots,\tilde{\phi}_{d_{M-1}},
  \tilde{\phi}_{d_{M+1}}]\bigr](x;\bm{\lambda})\\
  &\propto\text{W}\bigl[F(x,N,\bm{\lambda})
  \text{W}[\phi_0,\ldots,\breve{\phi}_{\bar{d}_1},\ldots,
  \breve{\phi}_{\bar{d}_{M-1}},\ldots,\breve{\phi}_{\bar{d}_M},\ldots,
  \phi_N],\\
  &\phantom{\propto\text{W}\bigl[}
  \ F(x,N,\bm{\lambda})
  \text{W}[\phi_0,\ldots,\breve{\phi}_{\bar{d}_1},\ldots,
  \breve{\phi}_{\bar{d}_{M-1}},\ldots,\breve{\phi}_{\bar{d}_{M+1}},\ldots,
  \phi_N]\bigr](x;\bar{\bm{\lambda}})\\
  &=F(x,N,\bm{\lambda})^2\text{W}\bigl[
  \text{W}[\phi_0,\ldots,\breve{\phi}_{\bar{d}_1},\ldots,
  \breve{\phi}_{\bar{d}_{M-1}},\ldots,\breve{\phi}_{\bar{d}_M},\ldots,
  \phi_N],\\
  &\phantom{=F(x,N,\bm{\lambda})^2\text{W}\bigl[}
  \ \text{W}[\phi_0,\ldots,\breve{\phi}_{\bar{d}_1},\ldots,
  \breve{\phi}_{\bar{d}_{M-1}},\ldots,\breve{\phi}_{\bar{d}_{M+1}},\ldots,
  \phi_N]\bigr](x;\bar{\bm{\lambda}})\\
  &=\pm F(x,N,\bm{\lambda})^2
  \text{W}[\phi_0,\ldots,\breve{\phi}_{\bar{d}_1},\ldots,
  \breve{\phi}_{\bar{d}_{M-1}},\ldots,
  \breve{\phi}_{\bar{d}_M},\ldots,
  \breve{\phi}_{\bar{d}_{M+1}},\ldots,\phi_N](x;\bar{\bm{\lambda}})\\
  &\phantom{=F(x,N,\bm{\lambda})^2}
  \times\text{W}[\phi_0,\ldots,\breve{\phi}_{\bar{d}_1},\ldots,
  \breve{\phi}_{\bar{d}_{M-1}},\ldots,\phi_N](x;\bar{\bm{\lambda}})\\
  &\propto F(x,N,\bm{\lambda})
  \text{W}[\phi_0,\ldots,\breve{\phi}_{\bar{d}_1},\ldots,
  \breve{\phi}_{\bar{d}_{M+1}},\ldots,\phi_N](x;\bar{\bm{\lambda}})\cdot
  \text{W}[\tilde{\phi}_{d_1},\ldots,\tilde{\phi}_{d_{M-1}}](x;\bm{\lambda}).
\end{align*}
This leads to
\begin{align*}
  &\quad\frac{\text{W}[\tilde{\phi}_{d_1},\ldots,\tilde{\phi}_{d_{M+1}}]
  (x;\bm{\lambda})}
  {A_{\{d_1,\ldots,d_{M+1}\}}(x;\bm{\lambda})}\n
  &\propto
  \frac{F(x,N,\bm{\lambda})
  \bar{A}_{\{0,\ldots,\bar{d}_1,\ldots,\bar{d}_{M+1},\ldots,N\}}
  (x;\bar{\bm{\lambda}})}
  {A_{\{d_1,\ldots,d_{M+1}\}}(x;\bm{\lambda})}
  \frac{\text{W}[\phi_0,\ldots,\breve{\phi}_{\bar{d}_1},\ldots,
  \breve{\phi}_{\bar{d}_{M+1}},\ldots,\phi_N](x;\bar{\bm{\lambda}})}
  {\bar{A}_{\{0,\ldots,\bar{d}_1,\ldots,\bar{d}_{M+1},\ldots,N\}}
  (x;\bar{\bm{\lambda}})}\n
  &=\frac{\text{W}[\phi_0,\ldots,\breve{\phi}_{\bar{d}_1},\ldots,
  \breve{\phi}_{\bar{d}_{M+1}},\ldots,\phi_N](x;\bar{\bm{\lambda}})}
  {\bar{A}_{\{0,\ldots,\bar{d}_1,\ldots,\bar{d}_{M+1},\ldots,N\}}
  (x;\bar{\bm{\lambda}})},
\end{align*}
which means
\begin{equation*}
  \Xi_{\{d_1,\ldots,d_{M+1}\}}(\eta;\bm{\lambda})
  \propto
  \bar{\Xi}_{\{0,\ldots,\bar{d}_1,\ldots,\bar{d}_{M+1},\ldots,N\}}
  (\eta;\bar{\bm{\lambda}}).
\end{equation*}
This concludes the induction proof of \eqref{detiden} and the proof of
Propositions \ref{main} and \ref{polywron} is completed.

\bigskip
In order to obtain the rational forms (the ratios of polynomials) of the
eigenfunctions \eqref{eigeq1}--\eqref{eigeq2} in Proposition \ref{main},
we need the Wronskian expressions of the numerators:
\begin{align}
  &\text{W}[\tilde{\phi}_{d_1},\tilde{\phi}_{d_2},\ldots,
  \tilde{\phi}_{d_M},\phi_n](x;\bm{\lambda})
  ={A}_{\mathcal{D},n}(x;\bm{\lambda})
  P_{\mathcal{D},n}\bigl(\eta(x);\bm{\lambda}\bigr),\n
  &A_{\mathcal{D},n}(x;\bm{\lambda})=\left\{
  \begin{array}{ll}
  \tilde{\phi}_{0(0)}(x;\bm{\lambda})^M
  \phi_0(x;\bm{\lambda})
  \bigl(\cF^{-1}\frac{d\eta(x)}{dx}\bigr)^{\frac12M(M-1)-M}
  &:\text{Group A}\\[4pt]
  \prod_{k=1}^M\phi_0\bigl(x;\mathfrak{t}(\bm{\lambda})+d_k\bm{\delta}\bigr)
  \cdot\phi_0(x;\bm{\lambda}+n\bm{\delta})
  &:\text{Group B}
  \end{array}\right..
\end{align}
Here $P_{\mathcal{D},n}(\eta;\bm{\lambda})$ is a polynomial in $\eta$
and its degree is generically $\ell_{\mathcal{D}}+M+n$.
This leads to
\begin{equation}
  \frac{\text{W}[\tilde{\phi}_{d_1},\ldots,\tilde{\phi}_{d_M},\phi_n]
  (x;\bm{\lambda})}
  {\text{W}[\tilde{\phi}_{d_1},\ldots,\tilde{\phi}_{d_M}](x;\bm{\lambda})}
  =\frac{P_{\mathcal{D},n}\bigl(\eta(x);\bm{\lambda}\bigr)}
  {\Xi_{\mathcal{D}}\bigl(\eta(x);\bm{\lambda}\bigr)}
  \times\left\{
  \begin{array}{ll}
  \phi_0(x;\bm{\lambda}-M\bm{\delta})
  &:\text{Group A}\\[2pt]
  \phi_0(x;\bm{\lambda}+n\bm{\delta})
  &:\text{Group B}
  \end{array}\right..
\end{equation}
Eq.\,\eqref{Wphi=AXi} leads to
\begin{align}
  &\frac{\text{W}[\phi_{d_1},\ldots,\phi_{d_M},\phi_n](x;\bm{\lambda})}
  {\text{W}[\phi_{d_1},\ldots,\phi_{d_M}](x;\bm{\lambda})}
  =\frac{\bar{P}_{\mathcal{D},n}\bigl(\eta(x);\bm{\lambda}\bigr)}
  {\bar{\Xi}_{\mathcal{D}}\bigl(\eta(x);\bm{\lambda}\bigr)}
  \times\left\{
  \begin{array}{ll}
  \phi_0(x;\bm{\lambda}+M\bm{\delta})
  &:\text{Group A}\\[2pt]
  \phi_0(x;\bm{\lambda}+n\bm{\delta})
  &:\text{Group B}
  \end{array}\right.,\n
  &\bar{P}_{\mathcal{D},n}(\eta;\bm{\lambda})\eqdef
  \bar{\Xi}_{\{d_1,\ldots,d_M,n\}}(\eta;\bm{\lambda})
  \ \ :\ \text{degree $=\ell_{\{d_1,\ldots,d_M,n\}}=~\ell_{\mathcal{D}}-M+n$}.
\end{align}
In the present case we have
\begin{equation*}
  \frac{\text{W}[\phi_{e_1},\ldots,\phi_{e_{N+1-M}},\phi_{N+1+n}]
  (x;\bar{\bm{\lambda}})}
  {\text{W}[\phi_{e_1},\ldots,\phi_{e_{N+1-M}}](x;\bar{\bm{\lambda}})}
  =\frac{\bar{P}_{\bar{\mathcal{D}},N+1+n}
  \bigl(\eta(x);\bar{\bm{\lambda}}\bigr)}
  {\bar{\Xi}_{\bar{\mathcal{D}}}\bigl(\eta(x);\bar{\bm{\lambda}}\bigr)}
  \times\left\{
  \begin{array}{ll}
  \phi_0(x;\bm{\lambda}-M\bm{\delta})
  &:\text{Group A}\\[2pt]
  \phi_0(x;\bm{\lambda}+n\bm{\delta})
  &:\text{Group B}
  \end{array}\right.,
\end{equation*}
where the degree of
$\bar{P}_{\bar{\mathcal{D}},N+1+n}(\eta;\bar{\bm{\lambda}})$ is
$\ell_{\bar{\mathcal{D}}}-(N+1-M)+(N+1+n)=\ell_{\mathcal{D}}+M+n$.
Therefore \eqref{eigeq1} implies
\begin{equation}
  P_{\mathcal{D},n}(\eta;\bm{\lambda})
  \propto\bar{P}_{\bar{\mathcal{D}},N+1+n}(\eta;\bar{\bm{\lambda}})
  \ \ (n\in\mathbb{Z}_{\geq 0}).
\end{equation}
Similarly \eqref{eigeq2} implies
\begin{equation}
  \Xi_{\{d_1,\ldots,\breve{d}_j,\ldots,d_M\}}(\eta;\bm{\lambda})
  \propto\bar{P}_{\bar{\mathcal{D}},\bar{d}_j}(\eta;\bar{\bm{\lambda}})
  \ \ (j=1,2,\ldots,M).
\end{equation}

Let us introduce $\mathcal{D}_\pm$ based on
$\mathcal{D}=\{d_1,d_2,\ldots,d_M\}$:
\begin{equation}
  \mathcal{D}_{\pm}=\{d_1\pm1,d_2\pm1,\ldots,d_M\pm1\}.
  \label{Dpm}
\end{equation}
The newly introduced polynomials at $n=0$ are related to the old ones:
\begin{align}
  &\bar{P}_{\mathcal{D},0}(\eta;\bm{\lambda})
  \propto
  \bar{\Xi}_{\mathcal{D}_-}(\eta;\bm{\lambda}+\bm{\delta}),
  \label{PoD0=}\\
  &P_{\mathcal{D},0}(\eta;\bm{\lambda})
  \propto
  \Xi_{\mathcal{D}_+}(\eta;\bm{\lambda}+\bm{\delta}).
  \label{PD0=}
\end{align}
Based on these formulas, the following relations follow:
\begin{align}
  &\bar{f}_{\mathcal{D}}(x;\bm{\lambda})
  \eqdef\log\biggl|
  \frac{\text{W}[\phi_{d_1},\ldots,\phi_{d_M},\phi_0](x;\bm{\lambda})}
  {\text{W}[\phi_{d_1},\ldots,\phi_{d_M}](x;\bm{\lambda})}\biggr|
  \quad(\min_j{d_j}\geq 2),\n
  &\quad\bigl(\partial_x\bar{f}_{\mathcal{D}}(x;\bm{\lambda})\bigr)^2
  -\partial_x^2\bar{f}_{\mathcal{D}}(x;\bm{\lambda})
  =\bigl(\partial_x\bar{f}_{\mathcal{D}_-}
  (x;\bm{\lambda}+\bm{\delta})\bigr)^2
  +\partial_x^2\bar{f}_{\mathcal{D}_-}(x;\bm{\lambda}+\bm{\delta})
  +\mathcal{E}_1(\bm{\lambda}),
  \label{fDsi-}\\
  &f_{\mathcal{D}}(x;\bm{\lambda})
  \eqdef\log\biggl|
  \frac{\text{W}[\tilde{\phi}_{d_1},\ldots,\tilde{\phi}_{d_M},\phi_0]
  (x;\bm{\lambda})}
  {\text{W}[\tilde{\phi}_{d_1},\ldots,\tilde{\phi}_{d_M}](x;\bm{\lambda})}
  \biggr|,\n
  &\quad\bigl(\partial_xf_{\mathcal{D}}(x;\bm{\lambda})\bigr)^2
  -\partial_x^2f_{\mathcal{D}}(x;\bm{\lambda})
  =\bigl(\partial_xf_{\mathcal{D}_+}(x;\bm{\lambda}+\bm{\delta})\bigr)^2
  +\partial_x^2f_{\mathcal{D}_+}(x;\bm{\lambda}+\bm{\delta})
  +\mathcal{E}_1(\bm{\lambda}),
  \label{fDsi+}
\end{align}
which have the same forms as the shape-invariance relation \eqref{shape2}
but they do not mean the shape-invariance.
The $M=1$ case for (M), (RM) and (Kh) was presented as `enlarged'
shape-invariance in \cite{quesne4,quesne5}.

In the rest of this section we provide the proofs of
\eqref{PoD0=}--\eqref{PD0=}.  
For \eqref{PoD0=}, we note that the forward shift relation \eqref{forward}
can be written as
\begin{equation}
  \frac{d}{dx}\frac{\phi_n(x;\bm{\lambda})}{\phi_0(x;\bm{\lambda})}
  =f_n(\bm{\lambda})
  \frac{\phi_{n-1}(x;\bm{\lambda}+\bm{\delta})}{\phi_0(x;\bm{\lambda})}.
\end{equation}
By using this, we obtain
\begin{align*}
  &\quad\text{W}[\phi_{d_1},\ldots,\phi_{d_M},\phi_0](x;\bm{\lambda})\n
  &=\phi_0(x;\bm{\lambda})^{M+1}
  \text{W}\Bigl[\frac{\phi_{d_1}}{\phi_0},\ldots,\frac{\phi_{d_M}}{\phi_0},1
  \Bigr](x;\bm{\lambda})\n
  &=\phi_0(x;\bm{\lambda})^{M+1}(-1)^M
  \text{W}\Bigl[\frac{d}{dx}\frac{\phi_{d_1}}{\phi_0},\ldots,
  \frac{d}{dx}\frac{\phi_{d_M}}{\phi_0}\Bigr](x;\bm{\lambda})\n
  &=\phi_0(x;\bm{\lambda})^{M+1}(-1)^M
  \text{W}\Bigl[f_{d_1}(\bm{\lambda})
  \frac{\phi_{d_1-1}(x;\bm{\lambda}+\bm{\delta})}{\phi_0(x;\bm{\lambda})},
  \ldots,f_{d_M}(\bm{\lambda})
  \frac{\phi_{d_M-1}(x;\bm{\lambda}+\bm{\delta})}{\phi_0(x;\bm{\lambda})}
  \Bigr](x)\n
  &=(-1)^M\prod_{j=1}^Mf_{d_j}(\bm{\lambda})\cdot
  \phi_0(x;\bm{\lambda})
  \text{W}[\phi_{d_1-1},\ldots,\phi_{d_M-1}](x;\bm{\lambda}+\bm{\delta}).
\end{align*}
This can be rewritten as
\begin{equation*}
  \frac{\bar{\Xi}_{\{d_1,\ldots,d_M,0\}}\bigl(\eta(x);\bm{\lambda}\bigr)}
  {\bar{\Xi}_{\{d_1-1,\ldots,d_M-1\}}
  \bigl(\eta(x);\bm{\lambda}+\bm{\delta}\bigr)}
  =(-1)^M\prod_{j=1}^Mf_{d_j}(\bm{\lambda})\cdot
  \frac{\phi_0(x;\bm{\lambda})
  \bar{A}_{\{d_1-1,\ldots,d_M-1\}}(x;\bm{\lambda}+\bm{\delta})}
  {\bar{A}_{\{d_1,\ldots,d_M,0\}}(x;\bm{\lambda})},
\end{equation*}
and the fact that its right hand side is a constant
\begin{equation*}
  \frac{\phi_0(x;\bm{\lambda})
  \bar{A}_{\{d_1-1,\ldots,d_M-1\}}(x;\bm{\lambda}+\bm{\delta})}
  {\bar{A}_{\{d_1,\ldots,d_M,0\}}(x;\bm{\lambda})}=1,
\end{equation*}
can be verified for any of the systems listed in
\S\,\ref{sec:H}--\S\,\ref{sec:hDPT}.
This concludes the proof of \eqref{PoD0=}.
We remark that \eqref{PoD0=} is
$\bar{\Xi}_{\{d_1,\ldots,d_M,0\}}(\eta;\bm{\lambda})\propto
\bar{\Xi}_{\{d_1-1,\ldots,d_M-1\}}(\eta;\bm{\lambda}+\bm{\delta})$,
which implies
\begin{equation}
  \bar{\Xi}_{\{0,1,\ldots,N\}}(\eta;\bm{\lambda})\propto
  \bar{\Xi}_{\{0,1,\ldots,N-1\}}(\eta;\bm{\lambda}+\bm{\delta})
  \propto\cdots\propto
  \bar{\Xi}_{\{0\}}(\eta;\bm{\lambda}+N\bm{\delta})
  =\text{constant}
  \label{Xi=const}.
\end{equation}

For \eqref{PD0=}, we note that the forward shift relation for
$\tilde{\phi}_{\text{v}}(x;\bm{\lambda})$ \eqref{twistfor}
can be rewritten as
\begin{equation}
  \frac{d}{dx}\frac{\tilde{\phi}_{\text{v}}(x;\bm{\lambda})}
  {\phi_0(x;\bm{\lambda})}
  =-\epsilon\,b_{\text{v}}(-\bm{\lambda})
  \frac{\tilde{\phi}_{\text{v}+1}(x;\bm{\lambda}+\bm{\delta})}
  {\phi_0(x;\bm{\lambda})}.
\end{equation}
By using this, we obtain
\begin{align*}
  &\quad\text{W}[\tilde{\phi}_{d_1},\ldots,\tilde{\phi}_{d_M},\phi_0]
  (x;\bm{\lambda})\n
  &=\phi_0(x;\bm{\lambda})^{M+1}
  \text{W}\Bigl[\frac{\tilde{\phi}_{d_1}}{\phi_0},\ldots,
  \frac{\tilde{\phi}_{d_M}}{\phi_0},1\Bigr](x;\bm{\lambda})\n
  &=\phi_0(x;\bm{\lambda})^{M+1}(-1)^M
  \text{W}\Bigl[\frac{d}{dx}\frac{\tilde{\phi}_{d_1}}{\phi_0},\ldots,
  \frac{d}{dx}\frac{\tilde{\phi}_{d_M}}{\phi_0}\Bigr](x;\bm{\lambda})\n
  &=\phi_0(x;\bm{\lambda})^{M+1}(-1)^M
  \text{W}\Bigl[-\epsilon\,b_{d_1}(-\bm{\lambda})
  \frac{\tilde{\phi}_{d_1+1}(x;\bm{\lambda}+\bm{\delta})}
  {\phi_0(x;\bm{\lambda})},
  \ldots,-\epsilon\,b_{d_M}(-\bm{\lambda})
  \frac{\tilde{\phi}_{d_M+1}(x;\bm{\lambda}+\bm{\delta})}
  {\phi_0(x;\bm{\lambda})}\Bigr](x)\n
  &=\prod_{j=1}^M\epsilon\,b_{d_j}(-\bm{\lambda})\cdot
  \phi_0(x;\bm{\lambda})
  \text{W}[\tilde{\phi}_{d_1+1},\ldots,\tilde{\phi}_{d_M+1}]
  (x;\bm{\lambda}+\bm{\delta}).
\end{align*}
This can be rewritten as
\begin{equation*}
  \frac{P_{\{d_1,\ldots,d_M\},0}\bigl(\eta(x);\bm{\lambda}\bigr)}
  {\Xi_{\{d_1+1,\ldots,d_M+1\}}\bigl(\eta(x);\bm{\lambda}+\bm{\delta}\bigr)}
  =\prod_{j=1}^M\epsilon\,b_{d_j}(-\bm{\lambda})\cdot
  \frac{\phi_0(x;\bm{\lambda})
  A_{\{d_1+1,\ldots,d_M+1\}}(x;\bm{\lambda}+\bm{\delta})}
  {A_{\{d_1,\ldots,d_M\},0}(x;\bm{\lambda})},
\end{equation*}
and the fact that its right hand side is a constant
\begin{equation*}
  \frac{\phi_0(x;\bm{\lambda})
  A_{\{d_1+1,\ldots,d_M+1\}}(x;\bm{\lambda}+\bm{\delta})}
  {A_{\{d_1,\ldots,d_M\},0}(x;\bm{\lambda})}=1,
\end{equation*}
can be verified for any of the systems listed in
\S\,\ref{sec:H}--\S\,\ref{sec:hDPT}.
This concludes the proof of \eqref{PD0=}.

\section{Summary and Comments}
\label{summary}

In the context of rational extensions of solvable potentials, the concept
of the pseudo virtual state wavefunctions is introduced.
They are obtained by relaxing two conditions of the virtual state
wavefunctions; the reciprocals are square-integrable at both boundaries
and they need not be nodeless.
A Darboux-Crum transformation in terms of a pseudo virtual state wavefunction
will produce a new eigenstate below the original ground state.
The same system can be derived by a special type of Krein-Adler transformation
with negatively shifted parameters.
The main results of the paper is the equivalence of Darboux-Crum
transformation in terms of multiple pseudo virtual states and Krein-Adler
transformations in terms of multiple eigenstates with negatively shifted
parameters.
This is based on polynomial Wronskian identities, which are generalisations
of those reported by the present authors \cite{gos} a few years ago.
The equivalence holds for most of the known shape-invariant potentials
consisting of eleven explicit examples having finite as well as infinite
discrete eigenstates.

The type $\II$ virtual state wavefunctions have been obtained by the discrete
symmetries for two potentials (C) \S\,\ref{sec:C} and (Kh) \S\,\ref{sec:Kh}.
They have been used in the context of $M=1$ rational extensions of these
potentials in \cite{grandati} and \cite{quesne5}.
Multi-indexed extensions of these two potentials can be constructed in
exactly the same way as in \cite{os25}.
In a separate publication \cite{os28}, we will discuss rational extensions
in terms of genuine virtual state wavefunctions for shape-invariant
potentials having finitely many discrete eigenstates.
They have different features from those having infinitely many eigenstates,
which have been explored in connection with the multi-indexed orthogonal
polynomials \cite{os25}.

The multi-indexed Laguerre and Jacobi orthogonal polynomials
are labeled by the multi-index $\mathcal{D}$, but different multi-index
sets may give the same multi-indexed polynomials, e.g. eqs.(50)--(51) in
\cite{os25}. The proposition \ref{polywron} gives its generalisation.
By applying the twist based on the type $\II$ discrete symmetry to
\eqref{detiden}, the l.h.s becomes the denominator polynomial with multiple
type $\I$ virtual state deletion and the r.h.s. becomes that of type $\II$.

After completing the manuscript, we became aware of a recent work
\cite{mar-quesne}, which discusses some rational extensions of the
harmonic oscillator. They correspond to some special examples of the
equivalence for the harmonic oscillator (H) and $M=2$.

\section*{Acknowledgements}
R.\,S. is supported in part by Grant-in-Aid for Scientific Research
from the Ministry of Education, Culture, Sports, Science and Technology
(MEXT), No.22540186.

\goodbreak

\goodbreak


\begin{thebibliography}{99}
%

\bibitem{os25}
S.\,Odake and R.\,Sasaki,
``Exactly Solvable Quantum Mechanics and Infinite Families of
Multi-indexed Orthogonal Polynomials,"
Phys. Lett. {\bf B702} (2011) 164-170,
{\tt arXiv:\hspace{0pt}1105.0508[math-ph]}.

\bibitem{gomez3}
D.\,G\'{o}mez-Ullate, N.\,Kamran and R.\,Milson,
``Two-step Darboux transformations and exceptional Laguerre polynomials,"
J. Math. Anal. Appl. {\bf 387} (2012) 410-418, 
{\tt arXiv:\hspace{0pt}1103.5724[math-ph]}.

\bibitem{genden}
L.\,E.\,Gendenshtein,
``Derivation of exact spectra of the Schroedinger equation by means of
supersymmetry,''
JETP Lett. {\bf 38} (1983) 356-359.

\bibitem{infhul}
L.\,Infeld and T.\,E.\,Hull,
``The factorization method,''
Rev. Mod. Phys. {\bf 23} (1951) 21-68.

\bibitem{susyqm}
F.\,Cooper, A.\,Khare and U.\,Sukhatme,
``Supersymmetry and quantum mechanics,''
Phys. Rep. {\bf 251} (1995) 267-385.

\bibitem{darb}
G.\,Darboux,
{\it Th\'eorie g\'en\'erale des surfaces}
vol 2 (1888) Gauthier-Villars, Paris.

\bibitem{crum}
M.\,M.\,Crum,
``Associated Sturm-Liouville systems,"
Quart. J. Math. Oxford Ser. (2) {\bf 6} (1955) 121-127,
{\tt arXiv:physics/9908019}.

\bibitem{dubov}
S.\,Yu.\,Dubov, V.\,M.\,Eleonski\u{i} and N.\,E.\,Kulagin,
``Equidistant spectra of anharmonic oscillators,''
 Soviet Phys. JETP {\bf 75} (1992) 446-451;
%
Chaos {\bf 4} (1994) 47-53.

\bibitem{gomez6}
D.\,G\'{o}mez-Ullate, N.\,Kamran and R.\,Milson,
``The Darboux transformation and algebraic deformations of shapeinvariant
potentials,"
J. Phys. {\bf A37} (2004) 1789-1804,
{\tt arXiv:quant-ph/0308062}.

\bibitem{gomez7}
D.\,G\'{o}mez-Ullate, N.\,Kamran and R.\,Milson,
``Supersymmetry and algebraic Darboux transformations,"
J. Phys. {\bf A37} (2004) 10065-10078,
{\tt arXiv:nlin.SI/0402052}.

\bibitem{grandati}
Y.\,Grandati, 
``Solvable rational extensions of the isotonic oscillator,''
Ann. Phys. {\bf 326} (2011) 2074-2090,
{\tt arXiv:1101.0055[math-ph]};
``Solvable rational extensions of the Morse and Kepler-Coulomb potentials,"
 J. Math. Phys. {\bf 52} (2011) 103505 (12pp),
{\tt arXiv:1103.5023[math-ph]};
``Multistep DBT and regular rational extensions of the isotonic oscillator,''
Ann. Phys. {\bf 327} (2012) 2411-2431,
{\tt arXiv:1108.4503[math-ph]};
``New rational extensions of solvable potentials with finite bound state
spectrum,''
Phys. Lett. {\bf A376} (2012) 2866-2872,
{\tt arXiv:1203.4149[math-ph]}.

\bibitem{ho2}
C.-L.\,Ho,
``Prepotential approach to solvable rational extensions of harmonic
oscillator and Morse potentials,"
J. Math. Phys. {\bf 52} (2011) 122107 (8pp), 
{\tt arXiv:1105.3670\hspace{0pt}[math-ph]}.

\bibitem{quesne4}
C.\,Quesne,
``Revisiting (quasi-)exactly solvable rational extensions of the Morse
potential,"
Int. J. Mod. Phys. {\bf A27} (2012) 1250073,
{\tt arXiv:1203.1812[math-ph]}.

\bibitem{quesne5}
C.\,Quesne,
``Novel Enlarged Shape Invariance Property and Exactly Solvable Rational
Extensions of the Rosen-Morse II and Eckart Potentials,"
SIGMA {\bf 8} (2012) 080 (19pp),
{\tt arXiv:1208.6165[math-ph]}.

\bibitem{adler}
M.\,G.\,Krein,
``On continuous analogue of a formula of Christoffel from the theory
of orthogonal polynomials," (Russian)
Doklady Acad. Nauk. CCCP, {\bf 113} (1957) 970-973;
V.\,\'E.\,Adler,
``A modification of Crum's method,''
Theor. Math. Phys. {\bf 101} (1994) 1381-1386.

\bibitem{gos}
L.\,Garc\'ia-Guti\'errez, S.\,Odake and R.\,Sasaki,
``Modification of Crum's Theorem for `Discrete' Quantum Mechanics,''
Prog. Theor. Phys. {\bf 124} (2010) 1-26,
{\tt arXiv:1004.0289\hspace{0pt}[math-ph]}.

\bibitem{schiff}
L.\,I.\,Schiff,
{\it Quantum mechanics},
third edition, McGraw-Hill (1968).

\bibitem{os7}
S.\,Odake and R.\,Sasaki,
``Unified theory of annihilation-creation operators for solvable
(`discrete') quantum mechanics,''
J. Math. Phys. {\bf 47} (2006) 102102 (33pp),
{\tt arXiv:\hspace{0pt}quant-ph/0605215};
``Exact solution in the Heisenberg picture and annihilation-creation
operators,"
Phys. Lett. {\bf B641} (2006) 112-117,
{\tt arXiv:quant-ph/0605221}.

\bibitem{os28}
S.\,Odake and R.\,Sasaki,
``Extensions of solvable potentials with finitely many discrete eigenstates,"
DPSU-12-4, YITP-12-96.

\bibitem{cal-ge}
F.\,Calogero and Y.\,Ge,
``Can the general solution of the second-order ODE characterizing Jacobi
polynomials be polynomial?"
J. Phys. {\bf A45} (2012) 095206.

\bibitem{mar-quesne}
I.\,Marquette and C.\,Quesne,
``Two-step rational extensions of the harmonic oscillator:
exceptional orthogonal polynomials and ladder operators,"
J. Phys. {\bf A46} (2013) 155201 (14pp), 
{\tt arXiv:1212.3474[math-ph]}.

\end{thebibliography}
\end{document}